\documentclass[10pt]{article}
\usepackage{graphicx}  
\usepackage[svgnames]{xcolor}
\usepackage{bbm}
\usepackage[authoryear]{natbib}
\usepackage{algorithm}
\usepackage{algpseudocode}  
\usepackage{appendix}
\usepackage[utf8]{inputenc}
\usepackage{amsmath}
\usepackage{amsfonts}
\usepackage{amssymb}
\usepackage{mathtools}  
\usepackage[version=4]{mhchem}
\usepackage{stmaryrd}
\usepackage{bbold}
\usepackage{enumitem}
\setlist[itemize]{label=\textbullet}
\usepackage{subcaption}
\usepackage{authblk}    
\usepackage[margin=0.8in]{geometry}
\usepackage{hyperref}
\usepackage{longtable}
\usepackage{booktabs}
\usepackage{multirow}
\usepackage{url}
\usepackage{setspace}  

\begin{document}
\title{Neural Posterior Estimation for Spatial Individual-Level Epidemic Models}
\author[1,*]{Yicheng Mao}
  \author[1,2]{Rob Deardon}
  \affil[1]{Department of Mathematics and Statistics, University of Calgary, University Drive NW, Calgary, T2N 1N4, Canada}
  \affil[2]{Faculty of Veterinary Medicine, University of Calgary, University Drive NW, Calgary, T2N 1N4, Canada}
  \affil[*]{Correspondence: yicheng.mao1@ucalgary.ca }

\date{}
\maketitle
\doublespacing 
\begin{abstract}
\noindent 
Spatial individual-level models (ILMs) provide a flexible framework for
modeling infectious disease transmission across populations with known
locations. Bayesian inference for these models relies on Markov chain
Monte Carlo (MCMC), which requires repeated likelihood evaluation and, when
parts of the epidemic trajectory are unobserved, data-augmented sampling
over high-dimensional latent variables. This computational cost limits
the applicability of MCMC to large populations and to settings requiring
inference across multiple outbreaks. We propose using neural posterior
estimation (NPE) for amortized Bayesian inference in spatial ILMs. NPE
trains a conditional normalizing flow on simulated data to approximate
the posterior directly, bypassing likelihood evaluation at inference
time. We compare two embedding architectures: a convolutional neural
network (CNN) operating on the population-level incidence curve and a graph
neural network (GNN) operating on individual-level infection and location data.
In a simulation study under full observation, stochastic removals, and
partial observation, both variants achieve empirical coverage close to the nominal level,
with the GNN embedding yielding lower error and narrower credible
intervals for the spatial transmission parameters. 
Data-augmented MCMC is the most accurate when its chains converge, and GNN-NPE is closer to MCMC than CNN-NPE at a small fraction of the per-epidemic cost of MCMC.
We apply the framework to the observed 2001 UK foot-and-mouth disease outbreak on 1,177 farms, where MCMC with the exact likelihood provides a reference posterior. Both NPE variants recover the reproduction number estimated by MCMC. GNN-NPE gives a narrower posterior for the spatial decay of transmission than CNN-NPE but is shifted relative to the MCMC reference.
\\
\textbf{Keywords:} simulation-based inference;
neural posterior estimation;
individual-level models;
spatial epidemic models;
foot-and-mouth disease
\end{abstract}


\section{Introduction}
\label{sec:intro}

Mathematical models of infectious disease transmission are central to
understanding epidemic dynamics and informing public health decisions.
Among the available modeling frameworks, individual-level models (ILMs)
represent each member of a population explicitly and allow infection
risk to depend on individual-level covariates such as spatial location,
susceptibility characteristics, and contact structure
\citep{deardon2010inference}. This flexibility makes ILMs applicable to
a range of epidemic settings where the spatial arrangement of
individuals influences transmission patterns.

Bayesian inference is a natural approach for estimating the parameters
of ILMs because it provides full posterior distributions that quantify
parameter uncertainty and support probabilistic forecasting. The standard
computational tool for Bayesian inference in ILMs is Markov chain Monte
Carlo (MCMC) \citep{almutiry2021epiilmct}. Inference becomes
computationally demanding as the population size and observation window
increase, because each likelihood evaluation requires repeated
calculation of individual infection probabilities over the evolving sets
of susceptible and infectious individuals. The cost increases further
when epidemic observations are incomplete. Infection times may be
censored, removal or recovery times may be unrecorded, and some
infections may go undetected entirely. In these settings, the unobserved
components of the epidemic trajectory must be sampled alongside the model
parameters using data-augmented MCMC
\citep{gibson1998estimating,oneill1999bayesian}, which produces a
high-dimensional augmented inference problem. This computational burden
is amplified in simulation-based evaluation, where parameter recovery,
coverage, and sensitivity analyses require inference across many
simulated datasets \citep{morris2019using}.

A range of computational methods have been developed to reduce the cost
of inference for spatial ILMs. One line of work accelerates
likelihood-based inference itself. Composite and cluster-based methods
replace parts of the full individual-level calculation with cheaper
approximations based on spatial groupings of individuals
\citep{ward2022Computationally,zhang2026Composite}.
These approaches lower the cost of each likelihood evaluation, but each new dataset still requires its own run of MCMC. 
A second
line of work replaces likelihood evaluation with simulation from the
model, an approach now formalized more broadly as simulation-based
inference (SBI) \citep{mckinley2014simulation,cranmer2020frontier}.
Approximate Bayesian computation (ABC) compares simulated and observed
data through summary statistics \citep{Minter2019Approximate} and has
been applied to spatial SEIR models for which MCMC becomes infeasible
\citep{Brown2018Approximate}. Gaussian process emulators construct fast
surrogates for model outputs and support emulation-based inference,
including settings with event-time uncertainty
\citep{pokharel2016gaussian,pokharel2022emulation}. ABC and
emulator-based methods target the posterior, but the inference procedure
must be run again for each new dataset. Supervised learning methods map
spatial epidemic data directly to predicted outcomes or transmission
classes \citep{pokharel2014supervised,augusta2019deep}. These methods
amortize the mapping from data to output across datasets, and they
return point predictions or class labels without a posterior
distribution. Thus, in earlier computational work on spatial ILMs,
posterior inference and amortization across datasets have largely been
developed separately.

Neural posterior estimation (NPE) is an SBI method that combines the
two \citep{papamakarios2018fast,Zammit-Mangion2025Neural}. It trains a conditional density estimator on simulated
parameter--data pairs to approximate the posterior distribution directly. Once trained, the
estimator returns a posterior for a new observation without likelihood
evaluation or latent-variable sampling, so the cost of simulation and
training is shared across all datasets analyzed with the same
estimator. NPE has been applied in astronomy
\citep{Dax2021real,Leyde2024Gravitational}, neuroscience
\citep{gonccalves2020training}, and economics \citep{dyer2024black}. In
epidemiology, NPE and related amortized methods have been applied to
compartmental models fitted to incidence, hospitalization, or intensive care time series
\citep{Radev2021OutbreakFlow,Wieland2026Assessment,bazarova2026simulation,Pinotti2025Simulation},
to pathogen phylogenies \citep{Voznica2022Deep,Pinotti2025Simulation},
to final outcome data from household-structured populations
\citep{kypraios2026neural}, to transmission among hospital patients
summarized by case counts \citep{chatha2024neural}, to multiple sites
that share global parameters \citep{Charles2026Tokenised}, and to
metapopulation models of national and subnational spread
\citep{bicker2026memiliohighperformance}. Several of these studies
compare NPE with particle or data-augmented MCMC and report posteriors
close to the likelihood-based reference at a fraction of the
computational cost
\citep{Wieland2026Assessment,bazarova2026simulation,kypraios2026neural}.
In most of these applications, the data passed to the estimator are
aggregated over individuals, as counts over time, final sizes, or counts by site.

Spatial ILMs record the infection status, timing, and location of each
individual, so information about the parameters lies in where infections
occur as well as in how many occur. Applying NPE to spatial ILMs
therefore raises a representation problem, namely how the embedding
network should reduce these data to a summary that retains this
information. The incidence curve is a natural choice for
population-level models, and convolutional neural networks (CNNs) can
extract temporal features from epidemic time series
\citep{lecun1998gradient,Lee2021predictive,Radev2021OutbreakFlow,Akesson2022Convolutional}.
Applied to a spatial ILM, the incidence curve discards the locations of
infected individuals, which carry information about the transmission
kernel. Graph neural networks (GNNs) can retain this information by
representing each individual as a node in a proximity graph, with node
features that combine infection status, timing, and location, and with
messages between neighbors weighted by distance
\citep{hamilton2017inductive}. Graph-based embeddings have been used for
amortized inference from irregularly located spatial data
\citep{SainsburyDale2025Neural}, phylogenetic trees
\citep{Leroy2025Graph}, individual-level output of agent-based models
\citep{dyer2022calibrating}, and epidemics on contact networks and
mobility graphs \citep{Backhausz2024Estimating,Wang2025Multi}, and their
use in epidemic modeling is reviewed in \citet{Liu2024Review}. In
agent-based models, adding spatial or temporal summaries does not always
improve neural posterior estimates \citep{Dignum2025Simulation}, so
whether the spatial information in ILMs improves posterior estimation
over the incidence curve alone has to be assessed empirically.

In this paper, we develop an NPE framework for Bayesian inference in
spatial ILMs and use it to make this assessment. We formulate NPE for
discrete-time spatial ILMs under three observation scenarios, namely
full observation, stochastic removals with latent removal times, and
partial observation with latent infection histories. In each scenario,
we compare a CNN embedding of the daily incidence curve with a GNN
embedding of the spatial graph of individuals. We benchmark both NPE
variants against likelihood-based MCMC in a controlled simulation study
and in an analysis of the observed 2001 UK foot-and-mouth disease (FMD)
outbreak. These comparisons assess whether the GNN embedding improves estimation of spatial transmission parameters relative to the CNN embedding, how closely NPE approximates the MCMC posterior, and at what computational cost.

The remainder of the paper is organized as follows.
Section~\ref{sec:model} defines the spatial ILM, the three observation
scenarios, and the data-augmented MCMC samplers used as benchmarks.
Section~\ref{sec:npe} describes the NPE framework, the two embedding
architectures, and the training procedure.
Section~\ref{sec:simulation} presents the simulation study.
Section~\ref{sec:fmd} describes the FMD application.
Section~\ref{sec:discussion} discusses the results and directions for
future work.


\section{Spatial SIR Individual-Level Models}
\label{sec:model}

\subsection{Spatial Individual-Level Models}

We consider a discrete-time SIR epidemic process
\citep{kermack1927contribution} within an individual-level modeling
framework. At each time step $t = 0,1,\ldots,T$, every individual is in
one of three disease states: susceptible ($S$), infectious ($I$), or
removed ($R$). Let $S_t$, $I_t$, and $R_t$ denote the sets of
susceptible, infectious, and removed individuals at time $t$,
respectively. Transitions follow the progression $S \to I \to R$. Once an individual enters the removed state, it no longer contributes to transmission.

Following \citet{deardon2010inference}, epidemic ILMs model infection
risk for each susceptible individual at each time step. The general
probability that susceptible individual $i$ becomes infectious at time
$t$ is
\begin{equation}
P^{SI}_{i,t} = 1 - \exp\!\left\{
-\left[
\Omega_S(i)\sum_{j \in I_t} \Omega_T(j)\,\kappa(i,j)
+ \epsilon_{i,t}
\right]
\right\},
\label{eq:ilm_general}
\end{equation}
where $\Omega_S(i) \geq 0$ represents factors affecting the
susceptibility of individual $i$, $\Omega_T(j) \geq 0$ captures
characteristics of infectious individual $j$ that influence
transmissibility, $\kappa(i,j) \geq 0$ is an infection kernel describing
the interaction between individuals $i$ and $j$, and
$\epsilon_{i,t} \geq 0$ is a spark term representing infection pressure
from sources not explicitly included in the model.

For the analyses in this paper, we use a parsimonious spatial ILM in
which susceptibility is governed by a common baseline parameter, all
infectious individuals are equally transmissible, and transmission
depends only on spatial proximity. Specifically, we set
$\Omega_S(i)=\alpha$, $\Omega_T(j)=1$,
$\kappa(i,j)=d_{ij}^{-\beta}$, and $\epsilon_{i,t}=0$, where $d_{ij}$
denotes the Euclidean distance between individuals $i$ and $j$. This
gives
\begin{equation}
P^{SI}_{i,t} = 1 - \exp\!\left(
-\alpha \sum_{j \in I_t} d_{ij}^{-\beta}
\right), \qquad \alpha,\beta > 0.
\label{eq:infection_prob}
\end{equation}
Here, $\alpha$ controls the overall intensity of transmission and
$\beta$ controls the rate at which infectious pressure decays with
distance. Larger values of $\alpha$ correspond to a more transmissible
epidemic under a fixed spatial configuration, whereas larger values of
$\beta$ imply that transmission is more strongly concentrated among
nearby individuals. Infection events for different susceptible
individuals at the same time step are assumed to be conditionally
independent given the current epidemic state.
Figure~\ref{fig:model}A illustrates a single time step of this process.

For removals, we consider two specifications. In the fixed-removal
specification, the infectious period is known and identical for all
individuals. In the stochastic-removal specification, each infectious
individual leaves the infectious compartment at each discrete time step with probability
\begin{equation}
P^{IR}=1-\exp(-\gamma),
\label{eq:removal_prob}
\end{equation}
where $\gamma>0$ is the removal-rate parameter. Thus, the epidemic
process parameter vector is $\theta=(\alpha,\beta)$ under fixed removals
and $\theta=(\alpha,\beta,\gamma)$ under stochastic removals.
The resulting state transitions are summarized in Figure~\ref{fig:model}B.

\subsection{Likelihood and Observation Scenarios}
\label{sec:likelihood}

Let $I^\star_{t+1}=S_t\setminus S_{t+1}$
denote the set of individuals newly infected between time $t$ and time
$t+1$. Conditional on the current epidemic state, newly infected
individuals contribute their infection probabilities, while individuals
who remain susceptible contribute their probabilities of avoiding
infection. The one-step contribution of the infection process is
\begin{equation}
  f_t\!\left(S_{t+1},I_{t+1},R_{t+1} \mid S_t,I_t,R_t,\theta\right)
  =
  \left[\prod_{i \in I^\star_{t+1}} P^{SI}_{i,t}\right]
  \left[\prod_{i \in S_{t+1}} \bigl(1-P^{SI}_{i,t}\bigr)\right],
  \label{eq:lik_t_fixed}
\end{equation}
where $P^{SI}_{i,t}$ is given by Eq.~\eqref{eq:infection_prob}. Under fixed
removals, removal times are deterministic given infection times, and the
likelihood of a complete epidemic trajectory $\mathcal{D}$ is
\begin{equation}
  f(\mathcal{D} \mid \theta)
  =
  \prod_{t=0}^{T-1}
  f_t\!\left(S_{t+1},I_{t+1},R_{t+1} \mid S_t,I_t,R_t,\theta\right).
  \label{eq:full_lik_fixed}
\end{equation}

Under stochastic removals, let $\tau_i$ and $r_i$ denote the infection
and removal times of infected individual $i$, respectively. The
infectious duration $g_i=r_i-\tau_i$ is the number of time steps during
which individual $i$ is infectious. A newly infected individual can be
removed in the same time step in which it becomes infected, before it
contributes to transmission, whereas the initial infectives are
infectious at $t=0$. Hence $g_i \geq \delta_i$, where $\delta_i=1$ for
the initial infectives and $\delta_i=0$ otherwise. Since removal occurs
with probability $P^{IR}$ at each discrete time step,
\begin{equation}
\Pr(g_i \mid \gamma)
=
(1-P^{IR})^{g_i-\delta_i}P^{IR},
\qquad g_i=\delta_i,\delta_i+1,\ldots.
\label{eq:duration_prob}
\end{equation}
For individuals still infectious at time $T$, $\Pr(g_i \mid \gamma)$ is replaced by
$(1-P^{IR})^{T-\tau_i-\delta_i}$.
For a complete trajectory with observed infection and removal times, the
likelihood is

\begin{equation}
  f(\mathcal{D} \mid \theta)
  =
  \prod_{t=0}^{T-1}
  f_t\!\left(S_{t+1},I_{t+1},R_{t+1} \mid S_t,I_t,R_t,\theta\right)
  \prod_{i:\tau_i<\infty} \Pr(g_i \mid \gamma).
  \label{eq:full_lik_stochastic_removal}
\end{equation}

We also consider partially observed epidemics. Let
$\mathcal{D}^{\mathrm{obs}}$ denote the observed infection data. In the
partial-observation setting, the population size $M$ and the locations
of all individuals are known, and each true infection is observed with
probability $\rho$, independently conditional on the complete latent
epidemic trajectory. An individual without an observed infection is therefore either uninfected or infected but unreported. Let $I_{t}^{\star,\mathrm{obs}}$ denote the observed
subset of the newly infected individuals $I_t^\star$. Then
\begin{equation}
  \Pr\!\left(I_{t}^{\star,\mathrm{obs}} \mid I_t^\star, \rho\right)
  =
  \rho^{|I_{t}^{\star,\mathrm{obs}}|}
  (1-\rho)^{|I_t^\star|-|I_{t}^{\star,\mathrm{obs}}|},
  \qquad I_{t}^{\star,\mathrm{obs}} \subseteq I_t^\star.
  \label{eq:obs_model}
\end{equation}
The complete-data likelihood factorizes as
\begin{equation}
  f(\mathcal{D},\mathcal{D}^{\mathrm{obs}} \mid \theta,\rho)
  =
  f(\mathcal{D} \mid \theta)
  f(\mathcal{D}^{\mathrm{obs}} \mid \mathcal{D},\rho),
  \label{eq:complete_partial_lik}
\end{equation}
where $f(\mathcal{D}\mid\theta)$ is given by
Eq.~\eqref{eq:full_lik_fixed} under fixed removals and by
Eq.~\eqref{eq:full_lik_stochastic_removal} under stochastic removals.

\begin{figure}[ht]
    \centering
    \includegraphics[width=\linewidth]{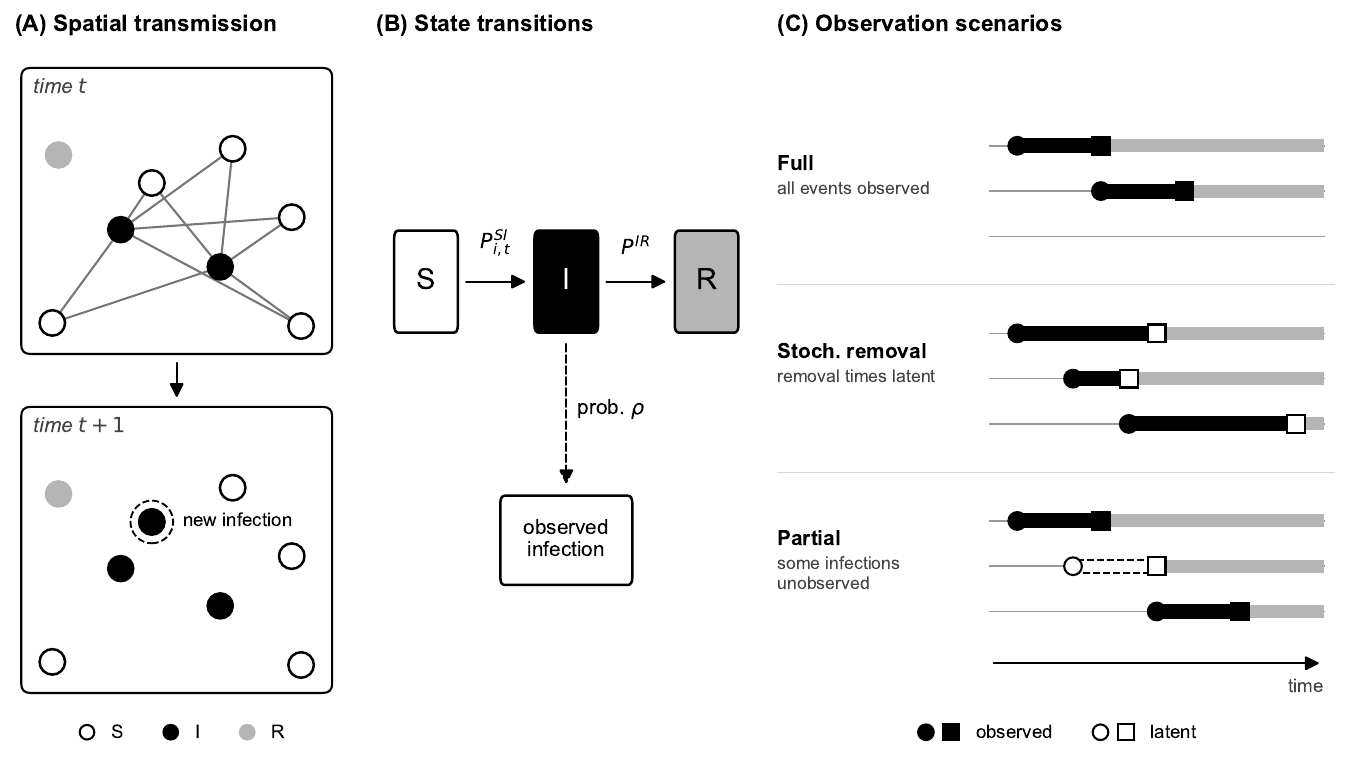}
    \caption{Spatial SIR individual-level model and observation scenarios.
    (A) Spatial transmission over one time step. At time $t$, every
    infectious individual exerts infectious pressure on every susceptible
    individual, with strength decreasing with distance; removed
    individuals exert no pressure. 
    Each susceptible individual is then
    infected independently with probability $P^{SI}_{i,t}$, giving the
    state at time $t+1$. (B) State transitions. In the partial observation
    scenario, each infection is observed with probability $\rho$.
    (C) Observation scenarios. Each row shows example event timelines for
    three individuals, with circles marking infection times and squares
    marking removal times. Filled symbols denote observed events and open
    symbols denote latent events; the dashed bar marks an infectious period
    that is never observed.}
    \label{fig:model}
\end{figure}

Combining the two removal specifications with the observation process
gives the three observation scenarios considered in this paper, illustrated
in Figure~\ref{fig:model}C. Under full observation (Full), removals are
fixed and all infection times are observed, so the complete trajectory is
available and inference targets $\theta$. Under stochastic removals
(Stoch.\ removal), infection times are observed but removal times are
latent, and inference again targets $\theta$. Under partial observation
(Partial), removals are fixed and infections are observed according to
Eq.~\eqref{eq:obs_model}, so the infection times of unobserved infected
individuals are latent and inference targets $(\theta,\rho)$.

\subsection{Bayesian Inference via Data-Augmented MCMC}
\label{sec:mcmc}

Under full observation, Bayesian inference is based on
\begin{equation}
  \pi(\theta \mid \mathcal{D}) \propto f(\mathcal{D} \mid \theta)p(\theta),
  \label{eq:posterior_full}
\end{equation}
where $p(\theta)$ is the prior distribution of the model parameters.
When removal times or infection histories are unobserved,
likelihood-based inference must account for the resulting latent
epidemic trajectory. A standard way to handle this problem is to use
data-augmented MCMC, in which the unobserved epidemic events are sampled
together with the model parameters \citep{almutiry2021epiilmct}.

In Stoch.\ removal, removal times are augmented as latent variables, and
the augmented posterior is
\begin{equation}
  \pi(\theta,\mathcal{D} \mid \mathcal{D}^{\mathrm{obs}})
  \propto
  f(\mathcal{D} \mid \theta)p(\theta),
  \label{eq:augmented_posterior_gamma}
\end{equation}
where $\mathcal{D}$ contains the observed infection times and augmented
removal times, and $f(\mathcal{D}\mid\theta)$ is given by
Eq.~\eqref{eq:full_lik_stochastic_removal}.

For partial observation, the observed data $\mathcal{D}^{\mathrm{obs}}$
contain only a subset of the true infection history. The augmented
posterior is
\begin{equation}
  \pi(\theta,\rho,\mathcal{D} \mid \mathcal{D}^{\mathrm{obs}})
  \propto
  f(\mathcal{D} \mid \theta)
  f(\mathcal{D}^{\mathrm{obs}} \mid \mathcal{D},\rho)
  p(\theta,\rho),
  \label{eq:augmented_posterior_partial}
\end{equation}
where $\mathcal{D}$ contains the augmented infection history, including
unobserved infection times.

We implemented the samplers in Python using NumPy
\citep{harris2020array} and SciPy \citep{virtanen2020scipy}. In all
three scenarios, $(\log\alpha,\beta)$ is updated jointly by an adaptive
random-walk Metropolis step \citep{roberts2009examples}, with adaptation
restricted to the burn-in period. Under Stoch.\ removal, each iteration
also updates $\gamma$ by a centered and a non-centered step
\citep{papaspiliopoulos2007general}, followed by 50 single-site
Metropolis--Hastings updates of the latent infectious durations, with
proposals drawn either from the duration model introduced in Eq.~\eqref{eq:duration_prob}
or by small shifts of the current duration. Under Partial, $\rho$ is
drawn from its Beta full conditional, and each iteration contains 200
Metropolis--Hastings updates that add, remove, or shift hidden infections, that is, infections of individuals that were not observed
to be infected. New hidden infections are proposed preferentially at
individuals and time steps with high current infection probability.
Details of all proposals and acceptance probabilities are given in
Section~S1 of the Supplementary Information.

Even when the epidemic trajectory is fully observed, repeated likelihood
evaluation can be computationally demanding in spatial ILMs because each
susceptible individual's infection probability depends on the combined
infectious pressure from all infectious individuals. This cost is
accumulated over all time points and repeated across many MCMC
iterations. The computational burden increases further when infection
times or removal times are augmented as latent variables. These
challenges motivate the use of simulation-based alternatives such as NPE,
which learns the conditional distribution of model parameters given
simulated epidemic observations without requiring repeated likelihood
evaluation at inference time.

\section{Neural Posterior Estimation for Spatial ILMs}
\label{sec:npe}

\subsection{Neural Posterior Estimation}
\label{sec:npe_framework}

NPE is a simulation-based inference approach that approximates the
Bayesian posterior $\pi(\vartheta \mid \mathcal{D}^{\mathrm{obs}})$ with
a parametric conditional density
$q_\phi(\vartheta \mid \mathcal{D}^{\mathrm{obs}})$ trained on simulated
data. Here, $\vartheta$ denotes the full parameter vector inferred in a
given observation scenario. For example, $\vartheta=\theta$ when only the
epidemic process parameters are inferred, while
$\vartheta=(\theta,\rho)$ in the partial-observation setting. We
parameterize $q_\phi(\vartheta \mid \mathcal{D}^{\mathrm{obs}})$ as a
normalizing flow conditioned on a learned summary of the observation:
\begin{equation}
  q_\phi(\vartheta \mid \mathcal{D}^{\mathrm{obs}})
  =
  q_\phi\!\bigl(\vartheta \mid h_\psi(\mathcal{D}^{\mathrm{obs}})\bigr),
  \label{eq:flow_with_embed}
\end{equation}
where $h_\psi : \mathcal{O} \to \mathbb{R}^{d_{\mathrm{emb}}}$ is an embedding network
that maps the observed epidemic data object to a fixed-length summary
vector, and $q_\phi(\cdot \mid h)$ is a normalizing flow conditioned on
$h$. The flow defines a flexible conditional density over the parameter
vector through an invertible transformation $T_\phi(\cdot \mid h)$ from
a standard base distribution $\pi_0$ to the target density:
\begin{equation}
  q_\phi(\vartheta \mid h)
  = \pi_0\!\left(T_\phi^{-1}(\vartheta \mid h)\right)
    \left|\det \partial T_\phi^{-1}/\partial \vartheta \right|.
  \label{eq:flow_density}
\end{equation}
Throughout this paper, $T_\phi$ is implemented as a masked autoregressive flow
\citep{papamakarios2017maf}.

Given the complete-trajectory simulator defined by
Eq.~\eqref{eq:infection_prob} and the prior $p(\vartheta)$, NPE constructs a
training set
$\{(\vartheta^{(n)}, \mathcal{D}^{\mathrm{obs},(n)})\}_{n=1}^{N}$. Each
training pair is generated by drawing parameters from the prior,
simulating a complete epidemic trajectory, and applying the observation
process that defines the data available for inference. The flow and
embedding parameters are estimated jointly by minimizing the expected
negative log-density
\begin{equation}
  \mathcal{L}(\phi, \psi)
  = -\mathbb{E}_{\vartheta \sim p(\vartheta),\,
      \mathcal{D}^{\mathrm{obs}} \sim
      p_{\mathrm{sim}}(\cdot \mid \vartheta)}
    \!\left[
    \log q_\phi\!\bigl(
    \vartheta \mid h_\psi(\mathcal{D}^{\mathrm{obs}})
    \bigr)
    \right],
  \label{eq:npe_loss}
\end{equation}
where $p_{\mathrm{sim}}(\mathcal{D}^{\mathrm{obs}}\mid\vartheta)$ is
the distribution over observed epidemic data induced by the complete
epidemic simulator and the corresponding observation process. The
population minimizer of Eq.~\eqref{eq:npe_loss} coincides with the true
posterior $\pi(\vartheta \mid \mathcal{D}^{\mathrm{obs}})$ for almost
every $\mathcal{D}^{\mathrm{obs}}$ under the simulator marginal
$p_{\mathrm{sim}}(\mathcal{D}^{\mathrm{obs}})$, provided the parametric
family $\{q_\phi \circ h_\psi\}$ is sufficiently expressive
\citep{papamakarios2021normalizing}.

The defining feature of NPE relative to likelihood-based methods is the
location of computational cost. Standard MCMC for spatial ILMs requires
repeated evaluation of the likelihood Eq.~\eqref{eq:full_lik_fixed} or
Eq.~\eqref{eq:full_lik_stochastic_removal} across many posterior samples.
NPE shifts this cost into a one-off simulation and training phase. Once
$q_\phi$ has been trained, posterior inference for a new observed
epidemic data object $\mathcal{D}^{\mathrm{obs}}$ reduces to sampling
from $q_\phi(\cdot \mid h_\psi(\mathcal{D}^{\mathrm{obs}}))$, without
repeated likelihood evaluation. The approach is amortized because the
same trained estimator can be evaluated on multiple observed datasets
generated under the same model.

\subsection{Embedding Architectures}
\label{sec:npe_embeddings}

\begin{figure}[h!]
    \centering
    \includegraphics[width=\linewidth]{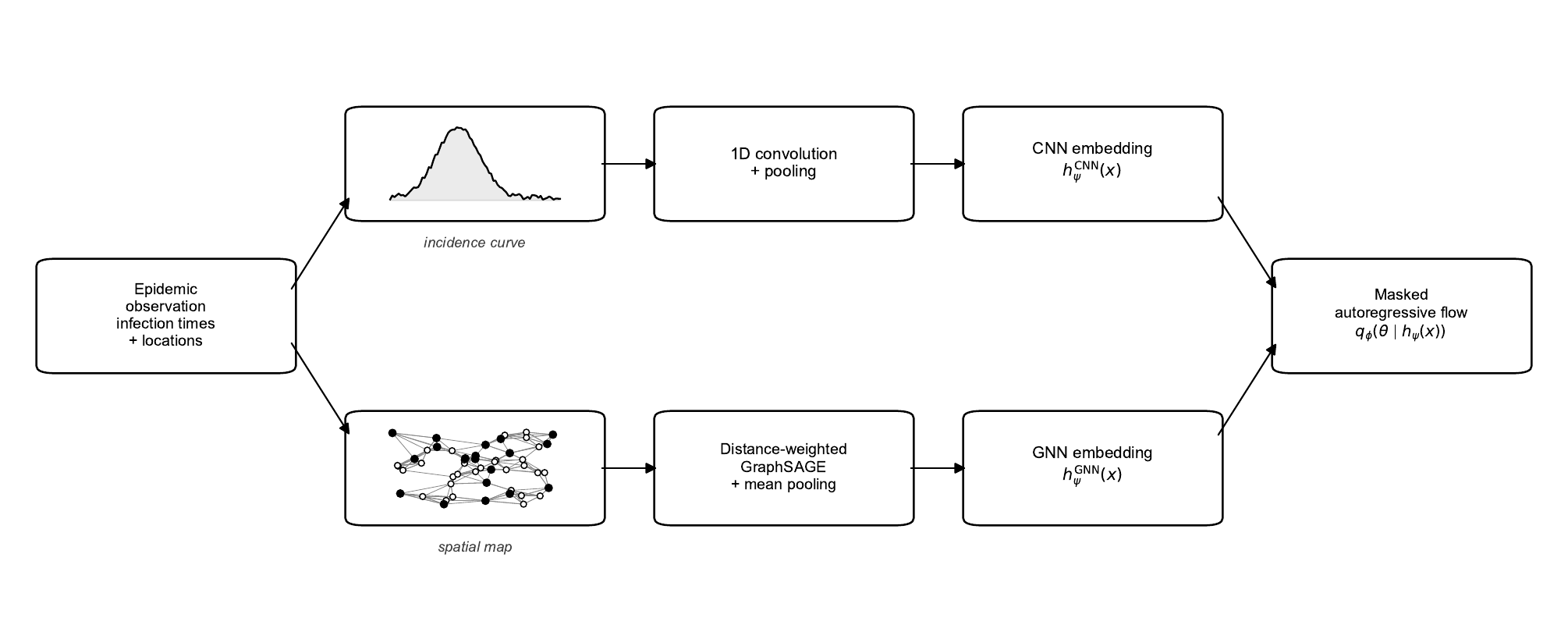}
    \caption{Schematic of the two embedding architectures used in this
    paper. The CNN branch summarizes the epidemic observation through a
    population-level observed incidence curve, whereas the GNN branch
    summarizes individual-level infection and location information
    through a spatial graph. Filled nodes in the graph illustration
    represent observed infections and open nodes represent individuals
    without observed infection. Both embeddings condition the same masked
autoregressive  flow used for posterior sampling.}
    \label{fig:embeddings}
\end{figure}

A single observed epidemic data object in our setting is represented by
infection-time encodings for a population of \(M\) individuals, together
with a fixed set of spatial coordinates. The infection-time encoding
uses a designated value for individuals without an observed infection,
so that observation status can be recovered directly from the encoding.
The embedding network \(h_\psi\) maps an architecture-specific
representation of these data to a fixed-length vector while retaining
information relevant for \(\vartheta\). Because \(h_\psi\) and
\(q_\phi\) are trained jointly under Eq.~\eqref{eq:npe_loss}, this reduction
is tailored to the posterior inference task. We compare two embedding
architectures that exploit different aspects of the observation, as
illustrated in Figure~\ref{fig:embeddings}.

\paragraph{Convolutional embedding (CNN).}
The CNN treats the epidemic as a temporal signal. From
$\mathcal{D}^{\mathrm{obs}}$, we form the daily observed incidence curve
  $\mathcal{C}
  =
  \left(I_1^{\star,\mathrm{obs}},\ldots,
  I_T^{\star,\mathrm{obs}}\right)^\top$. Because daily counts range over
several orders of magnitude across epidemics drawn from the prior, this
incidence curve is log-transformed and standardized before being passed
to the embedding network. A stack of one-dimensional convolution layers
with rectified linear unit (ReLU) activations \citep{nair2010rectified}
extracts local temporal features such as growth and decline over a few
days, and because the same filters are applied at every time step, these
features are detected wherever they occur in the epidemic. An adaptive
pooling layer then reduces the temporal dimension to a fixed number of
segments, which retains coarse information on when these features
occur, and a fully connected head projects the result to
$\mathbb{R}^{d_{\mathrm{emb}}}$. The CNN exploits the temporal structure
of the observed epidemic curve and is invariant to the labeling of
individuals, but discards the spatial coordinates.

\paragraph{Graph embedding (GNN).}
The GNN operates on a graph $G = (V, E)$ in which $V$ is the set of
individuals and $E$ contains directed edges from each individual $i$ to
its $k$ nearest spatial neighbors $\mathcal{N}_k(i)$. A sparse
nearest-neighbor graph keeps the cost of message passing linear in $M$,
and each layer extends the neighborhood that informs an individual by
one step, so stacked layers combine information over larger distances.
Each node carries a feature vector that combines an indicator of whether
the individual was observed to be infected, its normalized infection
time when observed, and its normalized spatial coordinates. These
features are first mapped to a hidden representation
$\mathbf{z}_i^{(0)}$ by a linear encoder and then passed through $L$
message-passing layers. Because edges defined by nearest-neighbor
relations alone do not distinguish a close neighbor from a distant one,
we use a distance-weighted variant of GraphSAGE
\citep{hamilton2017inductive}. Layer $l$ updates the representation of
individual $i$ as
\begin{equation}
  \mathbf{z}_i^{(l)}
  =
  \mathrm{ReLU}\!\left(
  \mathrm{LN}\!\left(
  \mathbf{A}^{(l)}\mathbf{z}_i^{(l-1)}
  + \mathbf{B}^{(l)}\sum_{j \in \mathcal{N}_k(i)} w_{ij}^{(l)}\mathbf{z}_j^{(l-1)}
  + \mathbf{c}^{(l)}
  \right)\right),
  \qquad
  w_{ij}^{(l)}
  =
  \frac{d_{ij}^{-\eta_l}}{\sum_{j' \in \mathcal{N}_k(i)} d_{ij'}^{-\eta_l}},
\end{equation}
where $\mathrm{LN}$ denotes layer normalization, $\mathbf{A}^{(l)}$,
$\mathbf{B}^{(l)}$ and $\mathbf{c}^{(l)}$ are learnable weights, and
$\eta_l$ is a learnable decay exponent for layer $l$. The neighbor
weights have the same power-law form as the spatial kernel in Eq.~\eqref{eq:infection_prob}. When $\eta_l = 0$ the weights are uniform
and the layer reduces to GraphSAGE with mean aggregation. We initialize
$\eta_l = 0$ so that the network starts from mean aggregation and
learns from the data how distance modulates the relative contributions
of neighboring individuals. A permutation-invariant readout by mean
pooling over $V$ and a fully connected head produce the final
embedding. Because the layer weights are shared across individuals, the
number of network parameters does not depend on $M$. The GNN preserves both the temporal and the spatial structure of the observation and is invariant
to relabeling of individuals. A comparison of this layer with mean aggregation is reported in
Section~S4 of the Supplementary Information.

The two architectures encode different inductive biases about which
features of the data are informative for $\vartheta$. The CNN summarizes
the population-level temporal pattern, whereas the GNN summarizes
individual-level information together with its spatial neighborhood.
Their comparative performance under the three observation scenarios is
discussed in Section~\ref{sec:simulation}.
Hyperparameters of both embedding networks and the flow were tuned with
Optuna \citep{akiba2019optuna}, and the search spaces and selected
configurations are reported in Section~S2 of the Supplementary Information.

\subsection{Training and Posterior Inference}
\label{sec:npe_training_inference}

Training NPE for the spatial ILM defined in Section~\ref{sec:model}
requires independent draws from the prior predictive distribution of the
observed epidemic data object used for inference. For each training pair,
a full parameter vector is drawn from its prior, a complete epidemic
trajectory is simulated from the forward model, and the simulated
trajectory is mapped to the data object available for inference. We write
this scenario-specific observation map as $g(\cdot;\vartheta)$:
\begin{equation}
  \vartheta^{(n)} \sim p(\vartheta), \qquad
  \mathcal{D}^{(n)} \sim f_D(\cdot \mid \theta^{(n)}), \qquad
  \mathcal{D}^{\mathrm{obs},(n)}
  = g(\mathcal{D}^{(n)};\vartheta^{(n)}),
  \label{eq:npe_training_pairs}
\end{equation}
where $\theta^{(n)}$ denotes the epidemic-process components of
$\vartheta^{(n)}$, and $f_D(\cdot \mid \theta^{(n)})$ denotes the
complete-trajectory simulator. The map $g(\cdot;\vartheta)$ encodes the
observation scenario and, when relevant, the observation process
parameters. Under full observation, $\mathcal{D}^{\mathrm{obs},(n)}$
contains the complete infection-time vector. Under stochastic removals,
removal times are generated as part of the epidemic trajectory, but
inference is based on the observed infection times. Under partial
observation, only a subset of true infections is retained and the
remaining infection times are treated as unobserved. Thus, the inputs
presented to the embedding network during training match the information
available to an analyst at inference time.

The embedding network and conditional density estimator are trained
jointly by minimizing the empirical version of Eq.~\eqref{eq:npe_loss}.
Because the training pairs are generated from the prior predictive
distribution induced by the same simulator and observation process, the
learned density targets the corresponding posterior distribution. In the
partially observed and stochastic-removal settings, NPE does not
explicitly sample latent infection or removal times at inference time.
Instead, the effect of these latent processes is learned from simulations
generated under the corresponding observation model.

Once $q_\phi$ has been trained, posterior samples for an observed
epidemic data object $\mathcal{D}^{\mathrm{obs}}$ are obtained by
computing $h_\psi(\mathcal{D}^{\mathrm{obs}})$ and drawing
\begin{equation}
  \vartheta^{(s)}
  \sim q_\phi(\cdot \mid h_\psi(\mathcal{D}^{\mathrm{obs}})),
  \qquad s=1,\ldots,S.
  \label{eq:npe_posterior_sampling}
\end{equation}
No likelihood evaluation
or latent-variable update is required at this stage. The complete
training and inference workflow is summarized in
Algorithm~\ref{alg:npe_spatial_ilm}.

NPE was implemented in Python using the sbi package
\citep{tejero-cantero2020sbi}. The CNN and GNN embedding networks,
including the GraphSAGE layers, were implemented directly in PyTorch
\citep{paszke2019pytorch}, and the $k$-nearest-neighbor graph was
precomputed with SciPy \citep{virtanen2020scipy}.

\begin{algorithm}[h]
\caption{Amortized NPE for spatial individual-level models}
\label{alg:npe_spatial_ilm}
\begin{algorithmic}[1]
\Require Prior $p(\vartheta)$, complete-trajectory simulator $f_D(\cdot \mid \theta)$, observation map $g(\cdot;\vartheta)$, embedding network $h_\psi$, conditional density estimator $q_\phi$
\Ensure Trained posterior estimator $q_\phi(\vartheta \mid h_\psi(\mathcal{D}^{\mathrm{obs}}))$

\State Initialize an empty training set $\mathcal{T}$.
\For{$n = 1,\ldots,N$}
    \State Draw parameters $\vartheta^{(n)} \sim p(\vartheta)$.
    \State Extract the epidemic-process components $\theta^{(n)}$ from $\vartheta^{(n)}$.
    \State Simulate a complete epidemic trajectory $\mathcal{D}^{(n)} \sim f_D(\cdot \mid \theta^{(n)})$.
    \State Construct the observed data object $\mathcal{D}^{\mathrm{obs},(n)} = g(\mathcal{D}^{(n)};\vartheta^{(n)})$.
    \State Add $(\vartheta^{(n)}, \mathcal{D}^{\mathrm{obs},(n)})$ to $\mathcal{T}$.
\EndFor
\State Train $h_\psi$ and $q_\phi$ on $\mathcal{T}$ by minimizing the empirical version of Eq.~\eqref{eq:npe_loss}.
\State For an observed data object $\mathcal{D}^{\mathrm{obs}}$, compute $h_\psi(\mathcal{D}^{\mathrm{obs}})$.
\State Draw posterior samples
$\vartheta^{(s)} \sim q_\phi(\cdot \mid h_\psi(\mathcal{D}^{\mathrm{obs}}))$ for $s=1,\ldots,S$.
\State Discard samples outside the prior support.
\end{algorithmic}
\end{algorithm}

\section{Simulation Study}
\label{sec:simulation}

We conducted a simulation study to evaluate NPE for spatial ILMs under
the three observation scenarios introduced in Section~\ref{sec:model}.
We evaluated whether graph-based embeddings improve NPE relative to incidence-based CNN embeddings and how both NPE approaches compare with likelihood-based MCMC.

\subsection{Simulation Setup}
\label{sec:simulation_setup}

All simulations used a spatial population of $M=500$ individuals located
uniformly in a $100 \times 100$ square region. The initial condition
included $I_0=3$ infective individuals and $M-I_0$ susceptible
individuals. Each epidemic was simulated for $T=40$ discrete time steps.
Transmission followed the spatial ILM in Eq.~\eqref{eq:infection_prob}.
Under fixed removals, the infectious period was set to three time steps.
Under stochastic removals, removal occurred according to the geometric
duration model induced by Eq.~\eqref{eq:removal_prob}. A schematic of the simulated spatial population is provided in Figure~S5 of the Supplementary Information.

We considered the three observation scenarios defined in
Section~\ref{sec:likelihood} and illustrated in Figure~\ref{fig:model}C.
The prior was specified through epidemiologically interpretable
quantities. In all scenarios, we assigned Gamma priors to the spatial decay parameter, $\beta \sim \mathrm{Gamma}(6,1/4)$, and the basic reproduction number, $R_0 \sim \mathrm{Gamma}(10,1/4)$, where Gamma distributions are parameterized by shape and scale. The transmission intensity $\alpha$ was then derived from $R_0$, $\beta$, and the spatial infectious pressure generated by the
initial infectives. Specifically, for a given value of $\beta$, let
$\mathcal{I}_0$ denote the set of initial infectives, with
$|\mathcal{I}_0|=I_0$, and define
\[
  \bar{\lambda}_0(\beta)
  =
  \frac{1}{|\mathcal{I}_0|}
  \sum_{j \in \mathcal{I}_0}
  \sum_{i \notin \mathcal{I}_0} d_{ij}^{-\beta}.
\]
This quantity denotes the average spatial infectious pressure generated
by the initial infectives. Under fixed removals, we set
\[
  \alpha
  =
  \frac{R_0}{\mu_{\mathrm{fix}}\bar{\lambda}_0(\beta)},
\]
where $\mu_{\mathrm{fix}}=3$ is the fixed infectious period. Under
stochastic removals, we used the same construction with
$\mu_I=1/P^{IR}$, the mean infectious duration of an initial infective
under Eq.~\eqref{eq:duration_prob}:
\[
  \alpha
  =
  \frac{R_0}{\mu_I\bar{\lambda}_0(\beta)}.
\]
This construction gave $R_0$ the same interpretation for the initial
infectives in the fixed- and stochastic-removal settings while allowing $\alpha$ to adjust
to the spatial configuration and distance-decay parameter.

Scenario-specific priors were added only for parameters introduced by
the observation or removal mechanism. In Partial, we used
$\rho\sim\mathrm{Uniform}(0,1)$ for the observation probability. In
Stoch.\ removal, we placed a comparatively concentrated log-normal prior
on $\mu_I$,
\[
  \mu_I \sim
  \mathrm{LogNormal}\!\left(\log(3)-\frac{0.27^2}{2},\,0.27\right),
\]
so that $E(\mu_I)=3$, and transformed it to the removal-rate parameter
through
\[
  \gamma=-\log\!\left(1-\frac{1}{\mu_I}\right).
\]
Individuals infected after time 0 can be removed in the time step in
which they are infected, so their mean infectious duration is
$\mu_I-1$. Further prior details and prior sensitivity analysis are provided in
Section~S5 of the Supplementary Information.

For each scenario, we generated independent training and test datasets by sampling from the prior predictive distribution induced by
$p(\vartheta)$, the forward simulator, and the scenario-specific observation process. All simulations shared the same individual locations and initial infectives, and each epidemic was generated from
its own parameter vector drawn from the prior. The final NPE analysis used 20,000 training simulations and 1,000 held-out test epidemics per
scenario. In a learning curve for Full (Figure~S1 in the Supplementary Information),
the MAE of GNN-NPE for
$\alpha$ and $\beta$ changed by less than 0.03 between 20,000 and 50,000 training simulations. For each scenario and embedding,
NPE was trained five times on the same training simulations with different random seeds, which change the network initialization, the
split into training and validation sets, and the minibatch order. Unless stated otherwise, NPE results are reported as the mean (SD)
over the five training runs. The CNN
and GNN posterior estimators were trained on the same simulated parameter
values and evaluated on the same test epidemics. Posterior inference used
3,000 posterior samples per test epidemic.

\subsection{Comparison of CNN and GNN Embeddings}
\label{sec:simulation_cnn_gnn}

We first compared the two embeddings on the 1,000 held-out test
epidemics of each scenario. Point estimation accuracy was assessed using mean absolute error (MAE)
of the posterior median. For test epidemic $n$ and parameter $m$, let
$\vartheta_{n,m}$ denote the true value and let
$\widehat{\vartheta}_{n,m}$ denote the posterior median. Then
\begin{equation}
  \mathrm{MAE}_m
  =
  \frac{1}{N_{\mathrm{test}}}
  \sum_{n=1}^{N_{\mathrm{test}}}
  \left|
  \widehat{\vartheta}_{n,m}
  -
  \vartheta_{n,m}
  \right|.
  \label{eq:mae}
\end{equation}
Posterior uncertainty was assessed using the empirical coverage and the
median width of the 95\% credible interval across test epidemics.
Calibration of the full posterior was assessed by simulation-based
calibration (SBC) \citep{talts2018validating}. For each test epidemic
and parameter, the SBC rank is the fraction of posterior draws that fall
below the true value. For a calibrated posterior estimator these ranks
are uniformly distributed on $[0,1]$, and departures of their histogram
from uniformity indicate bias or incorrect dispersion of the posterior.

Table~\ref{tab:cnn_gnn_simulation} compares CNN-NPE and GNN-NPE on the
full held-out test set. Across the three observation scenarios, GNN-NPE
improved estimation of the two transmission parameters, $\alpha$ and
$\beta$. In every scenario, GNN-NPE gave a lower MAE for both parameters
in all five training runs, and the reduction in mean MAE ranged from
42\% to 65\%. The improvement was clearest for $\beta$, the spatial decay
parameter. In Full, the MAE for $\beta$ decreased from 0.403 under
CNN-NPE to 0.141 under GNN-NPE. Similar reductions were observed in
Stoch.\ removal and Partial. GNN-NPE also produced narrower 95\%
credible intervals for $\alpha$ and $\beta$ while maintaining empirical
coverage close to the nominal level. 

The advantage of GNN-NPE was less consistent for parameters whose main
information source is not spatial. In Stoch.\ removal, CNN-NPE gave a
smaller MAE for $\gamma$ than GNN-NPE (0.078 and 0.090) with a narrower
credible interval, in all five training runs. In Partial, the two
embeddings produced similar MAE for $\rho$ (0.036 and 0.035), and
GNN-NPE produced a slightly narrower interval. These differences are
consistent with the inductive biases of the two embeddings. The
incidence curve directly represents temporal epidemic dynamics, whereas
the graph embedding is most informative for spatial transmission.
Empirical coverage was between 0.948 and 0.973 for both architectures in all three
scenarios.

\begin{figure}[h!]
    \centering
    \includegraphics[width=\linewidth]{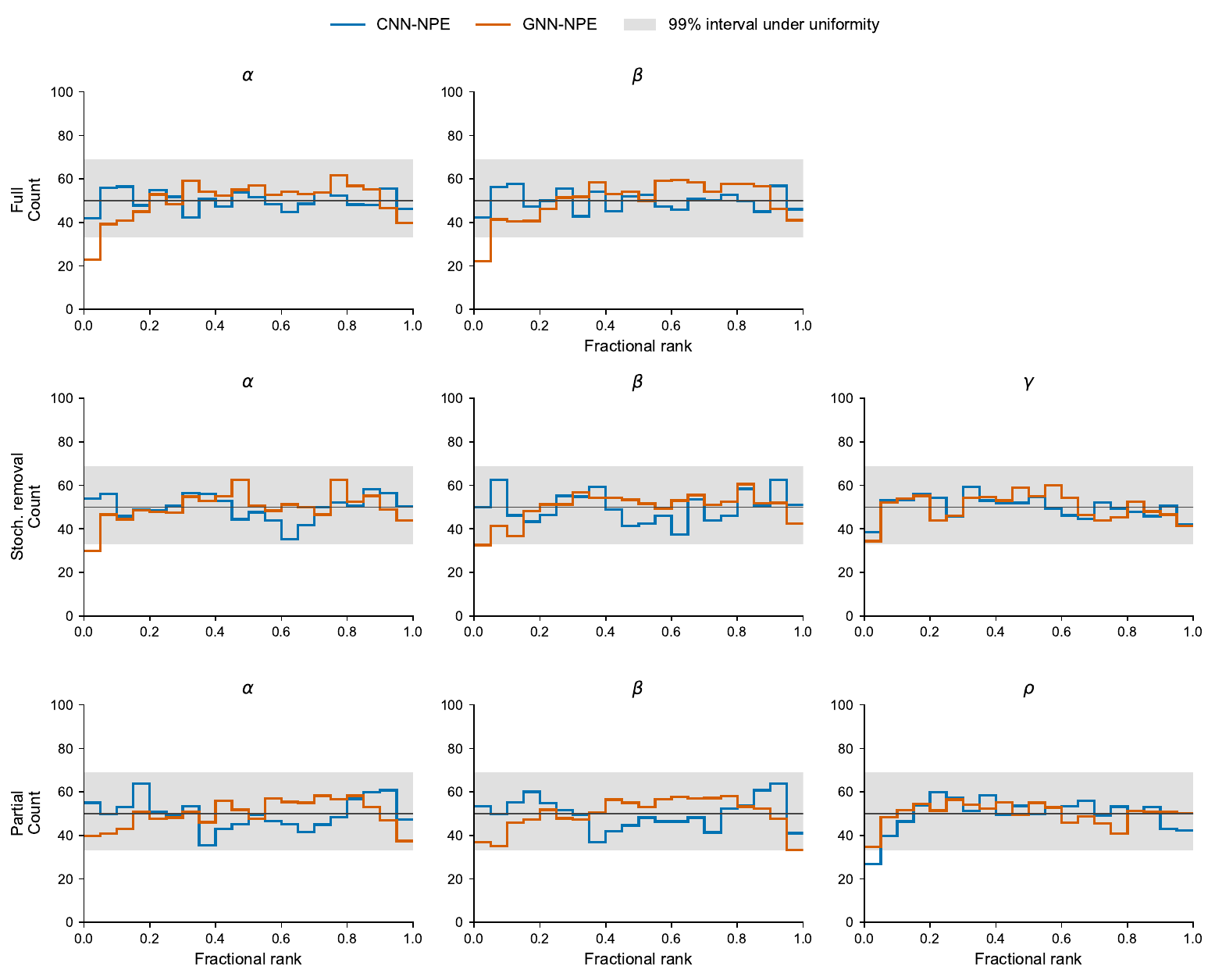}
    \caption{
    Simulation-based calibration rank histograms of CNN-NPE and GNN-NPE
    on the 1,000 held-out test epidemics per scenario. The rank of a
    test epidemic is the fraction of its 3,000 posterior draws that fall
    below the true value, and ranks are grouped into 20 bins. Counts are
    averaged over the five training runs. The gray band is the 99\%
    interval of the bin counts under uniform ranks, and the horizontal
    line is the expected count of 50.
    }
    \label{fig:sbc}
\end{figure}

Figure~\ref{fig:sbc} shows the SBC rank histograms on the 1,000 test
epidemics. A flat histogram indicates a calibrated posterior. A U-shaped
histogram indicates a posterior that is too narrow, an inverted U a
posterior that is too wide, and an excess at one end a posterior that is
shifted relative to the true value. The histograms of CNN-NPE were flat
and lay within the 99\% band expected under uniformity in every bin
except the lowest bin for $\rho$ in Partial. The histograms of GNN-NPE
had fewer ranks than expected in the extreme bins, most clearly in the
lowest bin, which fell below the band for $\alpha$ and $\beta$ in Full
and Stoch.\ removal. The deficit in the lowest bin means that the true
value fell below the 5th percentile of the posterior less often than
expected. The
largest deficit occurred in Full, with 22 to 23 ranks in the lowest bin
against 50 expected, and its size varied across the five training runs.
Together with coverage between 0.964 and 0.973 for the affected
parameters (Table~\ref{tab:cnn_gnn_simulation}), this pattern is
consistent with mildly conservative lower posterior tails. GNN-NPE
therefore gains its accuracy for $\alpha$ and $\beta$ without evidence
of undercoverage at the 95\% level, although the SBC histograms
indicate mild miscalibration in the posterior tails.

The comparison between the two embeddings was robust to the prior. In the sensitivity analysis described in Section~S5 of the Supplementary Information,
GNN-NPE gave a lower MAE than
CNN-NPE for $\alpha$ and $\beta$ under all seven prior specifications,
and the coverage of GNN-NPE remained between 0.942 and 0.967.

\begin{table}[h!]
\centering
\caption{
Performance of CNN-NPE and GNN-NPE on the full held-out test set of
1,000 epidemics. Entries are the mean (SD) over five independently
trained networks. MAE is the mean absolute error of the posterior
median. Width denotes the median width of the 95\% credible interval
across test epidemics. Coverage denotes empirical 95\% credible
interval coverage.
}
\label{tab:cnn_gnn_simulation}
\small
\setlength{\tabcolsep}{4pt}
\begin{tabular}{llccc ccc}
\toprule
 &  & \multicolumn{3}{c}{CNN-NPE} & \multicolumn{3}{c}{GNN-NPE} \\
\cmidrule(lr){3-5}\cmidrule(lr){6-8}
Scenario & Parameter & MAE & Width & Coverage & MAE & Width & Coverage \\
\midrule
Full & $\alpha$ & 0.675 (0.051) & 2.506 (0.792) & 0.959 (0.009) & 0.343 (0.020) & 0.405 (0.046) & 0.973 (0.007) \\
     & $\beta$  & 0.403 (0.033) & 1.821 (0.185) & 0.960 (0.010) & 0.141 (0.013) & 0.726 (0.047) & 0.971 (0.012) \\
\midrule
Stoch.\ removal & $\alpha$ & 0.652 (0.001) & 3.946 (0.312) & 0.948 (0.007) & 0.381 (0.009) & 0.734 (0.124) & 0.965 (0.007) \\
                & $\beta$  & 0.466 (0.001) & 2.076 (0.044) & 0.958 (0.006) & 0.219 (0.006) & 0.733 (0.056) & 0.964 (0.008) \\
                & $\gamma$ & 0.078 (0.000) & 0.353 (0.014) & 0.964 (0.006) & 0.090 (0.001) & 0.413 (0.019) & 0.963 (0.005) \\
\midrule
Partial & $\alpha$ & 0.684 (0.004) & 2.914 (0.360) & 0.959 (0.010) & 0.399 (0.015) & 0.839 (0.062) & 0.961 (0.010) \\
        & $\beta$  & 0.465 (0.002) & 1.922 (0.037) & 0.959 (0.003) & 0.226 (0.006) & 0.987 (0.023) & 0.967 (0.009) \\
        & $\rho$   & 0.036 (0.001) & 0.133 (0.006) & 0.969 (0.006) & 0.035 (0.001) & 0.121 (0.005) & 0.959 (0.010) \\
\bottomrule
\end{tabular}
\end{table}

\subsection{Comparison of NPE and MCMC}
\label{sec:simulation_npe_mcmc}

We further compared CNN-NPE and GNN-NPE with the data-augmented MCMC
samplers described in Section~\ref{sec:mcmc}. Because MCMC must be run
separately for each epidemic, we applied it to 100 epidemics sampled from
the 1,000 held-out test epidemics of each observation scenario. For each
epidemic, we ran four chains of 100,000 iterations and discarded the
first 20,000 iterations as burn-in. Convergence diagnostics were
computed from all 80,000 post-burn-in iterations of each chain. Every
10th iteration was stored, giving 8,000 draws per chain, and the
posterior summaries used 750 equally spaced stored draws per chain,
giving 3,000 draws per epidemic as for NPE.

Convergence was assessed with the rank-normalized split-$\hat{R}$ and
the bulk and tail effective sample sizes (ESS) \citep{vehtari2021rank},
computed for each parameter, the log-likelihood, and a summary of the
latent variables, namely the mean of the augmented infectious durations
in Stoch.\ removal
and the number of hidden infections in Partial. An epidemic was
classified as converged when all these quantities had $\hat{R}<1.01$
and bulk and tail ESS above 400. This criterion was met for all 100
epidemics in Full and Stoch.\ removal and for 65 of the 100 epidemics in
Partial (Tables~S6 and S7). In Partial, non-convergence was
concentrated at low observation probabilities and affected 26 of the 39
epidemics with $\rho<0.4$ and 9 of the 61 epidemics with $\rho\geq 0.4$
(Figure~S4A). The
comparison below uses the converged epidemics, and the non-converged runs in Partial are examined in Section S6 of the Supplementary Information.

\begin{table}[h!]
\centering
\caption{
CPU time of NPE and MCMC in the simulation study. NPE entries are the
mean (SD) over five training runs. 
Generation of the 20,000 training epidemics takes between 16 and 19
seconds per scenario and is excluded.
}
\label{tab:runtime_simulation}
\begin{tabular}{lccc}
\toprule
 & CNN-NPE & GNN-NPE & MCMC \\
\midrule
\multicolumn{4}{l}{\textit{One-off training (min)}} \\[2pt]
Full            & 9.6 (3.5) & 32.4 (10.5) & --- \\
Stoch.\ removal & 9.6 (2.0) & 23.8 (3.9)  & --- \\
Partial         & 9.2 (1.7) & 22.3 (2.2)  & --- \\
\midrule
\multicolumn{4}{l}{\textit{Inference per epidemic}} \\[2pt]
Full            & 0.056 (0.004)\,s & 0.138 (0.012)\,s & 11.0\,min \\
Stoch.\ removal & 0.087 (0.005)\,s & 0.263 (0.007)\,s & 26.5\,min \\
Partial         & 0.335 (0.008)\,s & 0.310 (0.086)\,s & 107.6\,min \\
\bottomrule
\end{tabular}
\end{table}

Table~\ref{tab:runtime_simulation} reports the computational cost of
each method. All computations were run on the CPU of a single Apple M5
Pro processor with 48\,GB of unified memory. NPE training is a one-off
cost of 9 to 10 minutes for CNN-NPE and 22 to 32 minutes for GNN-NPE per
scenario. A trained network returns 3,000 posterior draws in 0.06 to
0.34 seconds per epidemic. MCMC requires 11.0, 26.5, and 107.6 minutes per epidemic in Full, Stoch.\ removal, and Partial, respectively.
For applications that involve repeated inference across many observed datasets, the amortized structure of NPE provides a computational advantage that scales linearly with the number of inference targets.

\begin{table}[h!]
\centering
\caption{
Comparison of CNN-NPE, GNN-NPE, and MCMC on the epidemics with converged
MCMC (100 in Full, 100 in Stoch.\ removal, and 65 in Partial). MAE is
the mean absolute error of the posterior median. $r_S$ is the Spearman
correlation between true values and posterior medians. Width is the
median width of the 95\% credible interval. Coverage is the empirical
coverage of the 95\% credible interval. NPE entries are the mean (SD)
over five training runs.
}
\label{tab:npe_mcmc_simulation}
\small
\setlength{\tabcolsep}{4pt}
\begin{tabular}{lllcccc}
\toprule
Scenario & Method & Parameter & MAE & $r_S$ & Width & Coverage \\
\midrule
Full & CNN-NPE & $\alpha$ & 0.816 (0.068) & 0.304 (0.173) & 2.360 (0.888) & 0.958 (0.008) \\
     &         & $\beta$  & 0.401 (0.045) & 0.330 (0.101) & 1.781 (0.216) & 0.964 (0.009) \\
     & GNN-NPE & $\alpha$ & 0.456 (0.036) & 0.914 (0.013) & 0.282 (0.033) & 0.962 (0.008) \\
     &         & $\beta$  & 0.145 (0.012) & 0.916 (0.013) & 0.761 (0.051) & 0.956 (0.011) \\
     & MCMC    & $\alpha$ & 0.282 & 0.970 & 0.113 & 0.900 \\
     &         & $\beta$  & 0.073 & 0.973 & 0.269 & 0.920 \\
\midrule
Stoch.\ removal
     & CNN-NPE & $\alpha$ & 0.574 (0.002) & 0.303 (0.030) & 3.404 (0.234) & 0.938 (0.008) \\
     &         & $\beta$  & 0.432 (0.004) & 0.143 (0.076) & 1.990 (0.050) & 0.962 (0.004) \\
     &         & $\gamma$ & 0.081 (0.001) & 0.787 (0.007) & 0.320 (0.015) & 0.990 (0.000) \\
     & GNN-NPE & $\alpha$ & 0.321 (0.012) & 0.837 (0.015) & 0.753 (0.104) & 0.972 (0.013) \\
     &         & $\beta$  & 0.195 (0.005) & 0.840 (0.012) & 0.717 (0.056) & 0.966 (0.023) \\
     &         & $\gamma$ & 0.097 (0.003) & 0.691 (0.014) & 0.381 (0.018) & 0.958 (0.015) \\
     & MCMC    & $\alpha$ & 0.194 & 0.940 & 0.297 & 0.970 \\
     &         & $\beta$  & 0.118 & 0.932 & 0.349 & 0.940 \\
     &         & $\gamma$ & 0.067 & 0.835 & 0.242 & 0.940 \\
\midrule
Partial
     & CNN-NPE & $\alpha$ & 0.258 (0.005) & 0.157 (0.048) & 2.778 (0.224) & 0.948 (0.018) \\
     &         & $\beta$  & 0.390 (0.005) & 0.026 (0.034) & 1.843 (0.011) & 0.951 (0.007) \\
     &         & $\rho$   & 0.028 (0.002) & 0.960 (0.005) & 0.139 (0.007) & 0.966 (0.017) \\
     & GNN-NPE & $\alpha$ & 0.119 (0.009) & 0.859 (0.003) & 0.371 (0.033) & 0.914 (0.018) \\
     &         & $\beta$  & 0.177 (0.007) & 0.865 (0.008) & 0.855 (0.022) & 0.935 (0.017) \\
     &         & $\rho$   & 0.030 (0.001) & 0.957 (0.002) & 0.122 (0.006) & 0.938 (0.022) \\
     & MCMC    & $\alpha$ & 0.057 & 0.957 & 0.171 & 0.954 \\
     &         & $\beta$  & 0.102 & 0.961 & 0.477 & 0.954 \\
     &         & $\rho$   & 0.024 & 0.965 & 0.093 & 0.954 \\
\bottomrule
\end{tabular}
\end{table}

Table~\ref{tab:npe_mcmc_simulation} compares the three methods on the
converged epidemics. In addition to MAE, width, and coverage, the table
reports the Spearman correlation $r_S$ between true values and posterior
medians across epidemics, which measures how well a method orders
epidemics by their parameter values. MCMC gave the lowest MAE and the highest $r_S$ for
$\alpha$ and $\beta$ in all three scenarios, with 95\% credible
intervals between 1.8 and 2.8 times narrower than those of GNN-NPE and
coverage between 0.900 and 0.970. GNN-NPE was closest to MCMC, with
$r_S$ between 0.837 and 0.916 for $\alpha$ and $\beta$ and coverage
between 0.914 and 0.972. CNN-NPE had $r_S$ between 0.026 and 0.330 for
$\alpha$ and $\beta$, so its posterior medians carried little
information about the ordering of epidemics by their transmission
parameters, and its intervals for $\alpha$ were 4.5 to 8.4 times wider
than those of GNN-NPE. For the parameters whose main information source
is temporal, the three methods performed similarly. For $\gamma$, MCMC
had the lowest MAE (0.067), followed by CNN-NPE (0.081) and GNN-NPE
(0.097). For $\rho$, MAE ranged from 0.024 to 0.030 and $r_S$ from 0.957
to 0.965 across methods. MCMC uses the likelihood of the complete or
augmented data, and its advantage on the converged epidemics is
consistent with information lost when an epidemic is compressed into a
learned embedding and with approximation error of the network.  

\begin{figure}[h!]
    \centering
    \includegraphics[width=\linewidth]{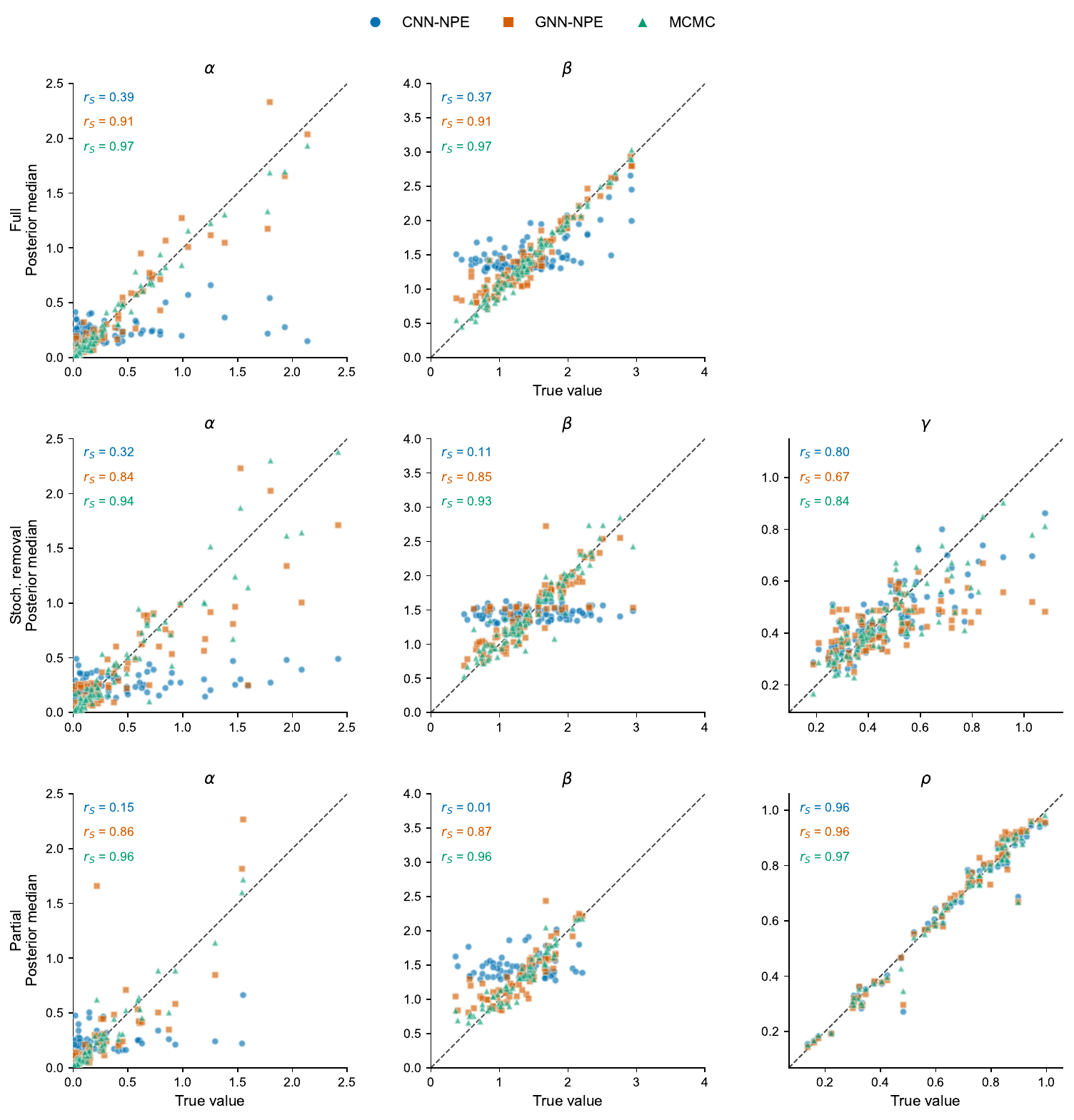}
    \caption{
    Parameter recovery on the epidemics with converged MCMC. Points
    compare true parameter values with posterior medians for CNN-NPE,
    GNN-NPE, and MCMC, with NPE results from the first of the five
    training runs. The dashed line indicates exact recovery, and $r_S$ is
    the Spearman correlation between true values and posterior medians
    for each method.
    }
    \label{fig:npe_mcmc_recovery}
\end{figure}

Figure~\ref{fig:npe_mcmc_recovery} plots posterior medians against true
values. The MCMC medians lie close to the diagonal for all parameters.
GNN-NPE medians follow the diagonal with more scatter, and for both
methods the scatter increases with the true value of $\alpha$. CNN-NPE
medians form a horizontal band across the prior range, at about 0.1 to
0.6 for $\alpha$ and at about 1.3 to 1.7 for $\beta$, around the prior
mean of 1.5, which explains the low $r_S$ in
Table~\ref{tab:npe_mcmc_simulation}. For $\rho$, the medians of all
three methods lie on the diagonal.

\begin{figure}[h!]
    \centering
    \includegraphics[width=\linewidth]{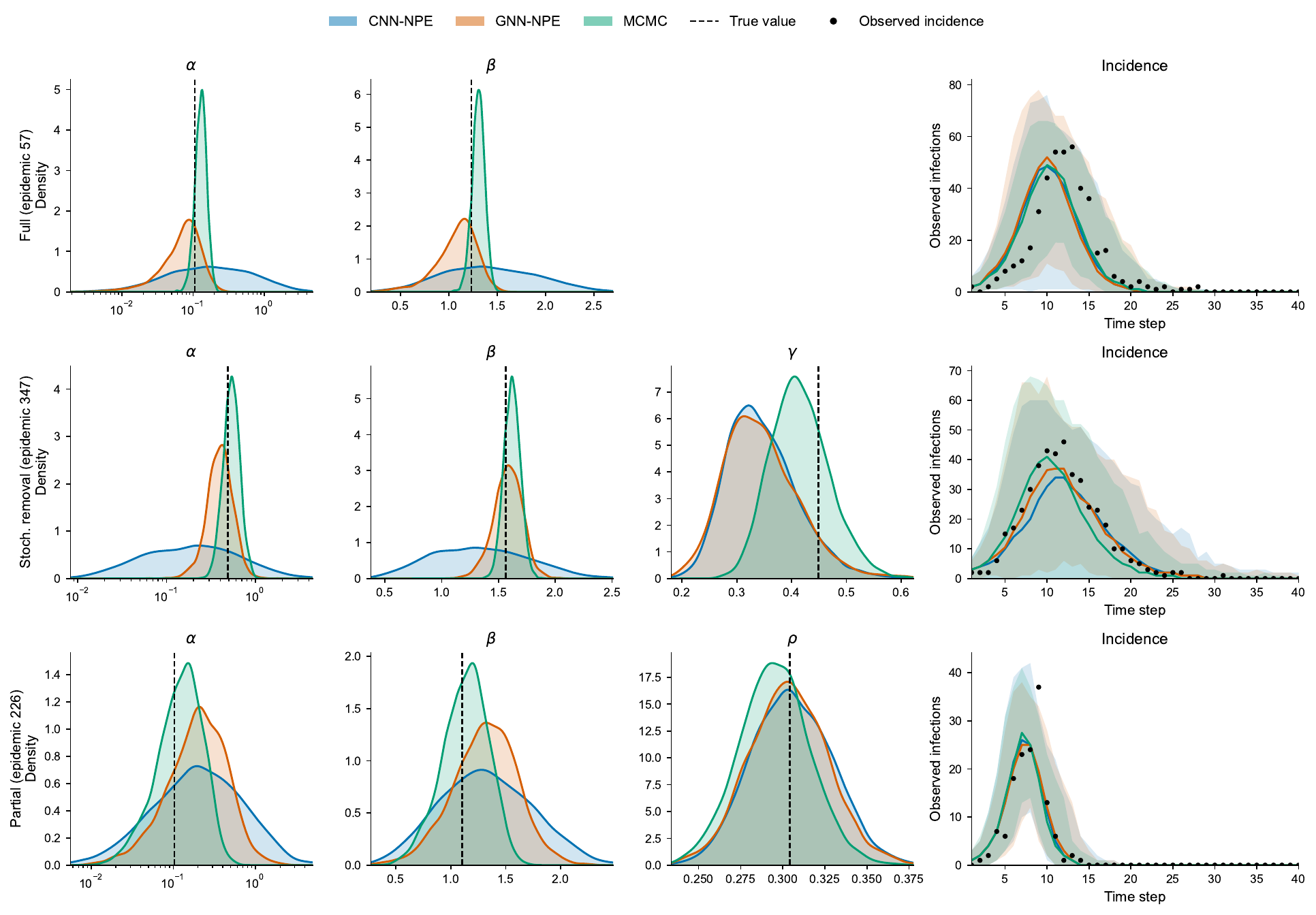}
    \caption{
    Posterior distributions and posterior predictive incidence for one
    epidemic per scenario. Each epidemic was selected among the epidemics
    with converged MCMC, at least 50 observed infections, and MCMC 95\%
    credible intervals that contain all true parameter values, as the
    epidemic whose mean rank of GNN-NPE absolute errors across parameters
    is closest to the median. The first three columns show kernel density
    estimates of the marginal posteriors, with $\alpha$ on a log scale
    and the true values as dashed lines. The last column shows the
    pointwise median (line) and 95\% interval (shaded) of the daily
    observed incidence simulated from 200 posterior draws per method,
    with the observed incidence as points. NPE results are from the
    first training run.
    }
    \label{fig:npe_mcmc_individual}
\end{figure}

We also assessed model fit through posterior predictive checks.
For each epidemic and each inference method, parameter vectors were
drawn from the estimated posterior and used to simulate new epidemic
trajectories on the same spatial population using the forward simulator
described in Section~\ref{sec:model}. The resulting predictive incidence
curves were summarized by their pointwise median and 95\% posterior
predictive interval and compared with the observed incidence curve.
Figure~\ref{fig:npe_mcmc_individual} shows the posteriors and posterior
predictive incidence for one representative epidemic per scenario. In
each epidemic, the MCMC posterior is the most concentrated, the GNN-NPE
posteriors for $\alpha$ and $\beta$ overlap it with a larger spread, and
the CNN-NPE posterior for $\alpha$ extends over two to three orders of
magnitude. For $\gamma$ in Stoch.\ removal, the two NPE posteriors
coincide and place the true value in their upper tail. For $\rho$ in
Partial, the three posteriors nearly coincide. The 95\% posterior
predictive intervals of all three methods contain the observed daily
incidence at almost all time steps. Averaged over the converged
epidemics, the posterior predictive medians of all three methods follow
the mean observed incidence curve in every scenario (Figure~S6).

\begin{figure}[h!]
    \centering
    \includegraphics[width=\linewidth]{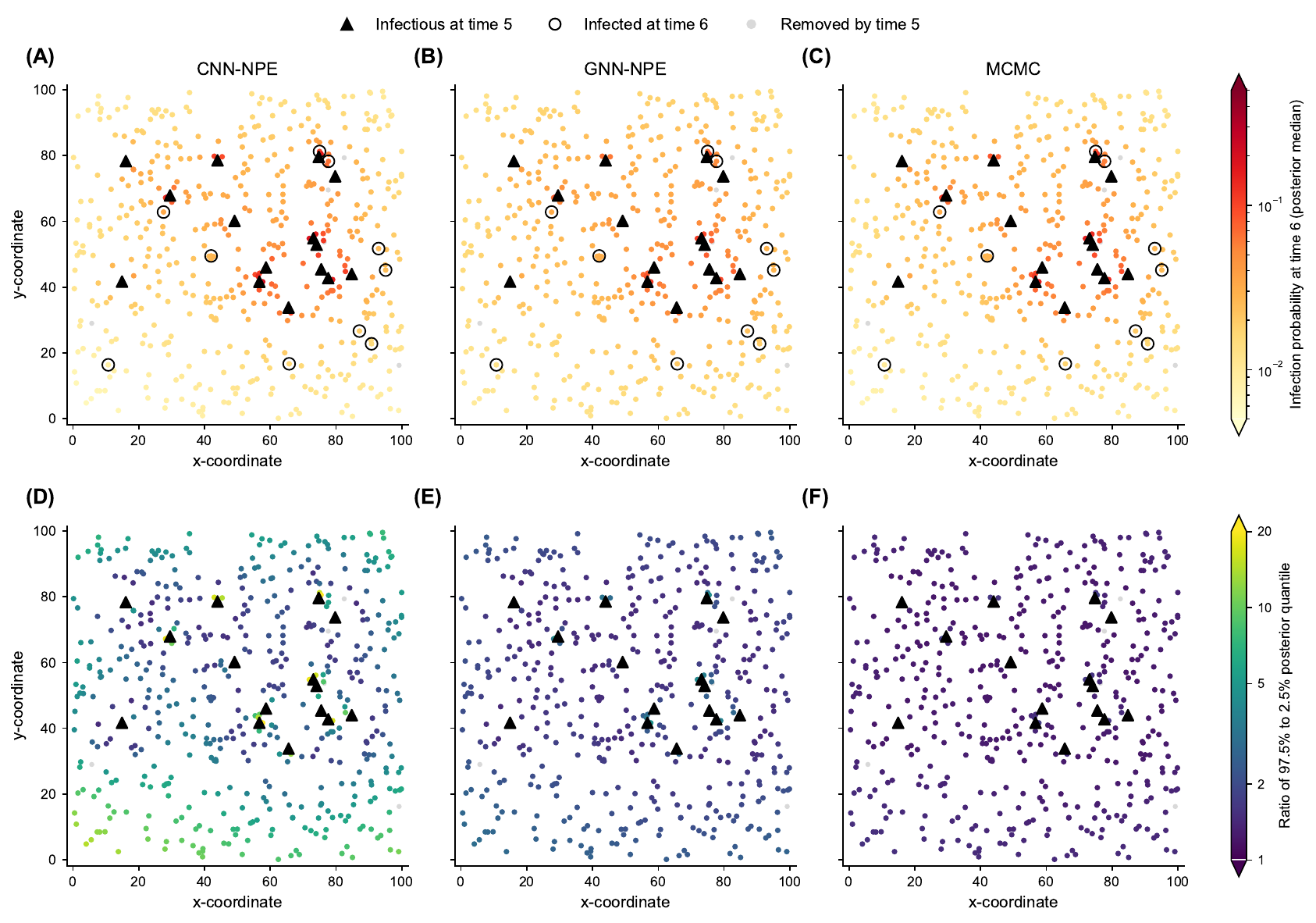}
    \caption{
    One-step-ahead infection probabilities for the Full epidemic in
    Figure~\ref{fig:npe_mcmc_individual}. Panels A to C show the
    posterior median of the probability that each susceptible
    individual is infected at time 6 given the epidemic state at time 5.
    Panels D to F show the ratio of the 97.5\% to the 2.5\% posterior
    quantile of this probability. Probabilities were computed from 1,000
    posterior draws per method. Triangles mark individuals infectious at
    time 5, open circles mark individuals infected at time 6, and gray
    points mark individuals removed by time 5.
    }
    \label{fig:sim_risk}
\end{figure}

Figure~\ref{fig:sim_risk} examines model fit at the level of
individuals, using the Full epidemic shown in
Figure~\ref{fig:npe_mcmc_individual}. We took the epidemic state at
time 5, when
15 individuals were infectious and 480 were susceptible. For each
susceptible individual, we evaluated its probability of infection at
time 6 under Eq.~\eqref{eq:infection_prob} for 1,000 posterior draws of each
method, which gives a posterior distribution of this probability for
every susceptible individual. Panels A to C map the posterior median of
the probability, and panels D to F map the ratio of its 97.5\% to its
2.5\% posterior quantile, which measures how uncertain the probability
is. The median maps of the three methods are similar, with the highest
probabilities near the infectious individuals and lower probabilities
further away.
The ratio maps differ between the
methods. The median ratio across susceptible individuals was 3.46 for
CNN-NPE, 1.77 for GNN-NPE, and 1.27 for MCMC. Under CNN-NPE, the ratio
increases with distance from the infectious individuals and exceeds 10
in the lower left of the region. This pattern is consistent with the weak
recovery of $\beta$ by CNN-NPE (Table~\ref{tab:npe_mcmc_simulation}). The infectious pressure from an
individual at distance $d$ is proportional to $d^{-\beta}$, so an error
in $\beta$ changes the infection probability by a factor that grows
with $d$.

Among the converged epidemics, MCMC gave the most accurate posteriors
at a cost of 11 to 108 minutes per epidemic. GNN-NPE was closer to
MCMC than CNN-NPE for $\alpha$ and $\beta$, with 95\% intervals 1.8 to
2.8 times wider than those of MCMC, and returned each posterior in less
than a second after training. When many epidemics must be analyzed or
posteriors are needed quickly, this difference in cost favors GNN-NPE.
The convergence failures of MCMC in Partial at low observation
probabilities illustrate a setting in which an amortized estimator may
be especially useful.

\section{Application to the 2001 UK Foot-and-Mouth Disease Outbreak}
\label{sec:fmd}

We applied both NPE variants to farm-level data from the 2001 UK
foot-and-mouth disease (FMD) outbreak and compared them with MCMC under
the same model.

\subsection{Data and Model}
\label{sec:fmd_data}

\begin{figure}[h!]
    \centering
    \includegraphics[width=\linewidth]{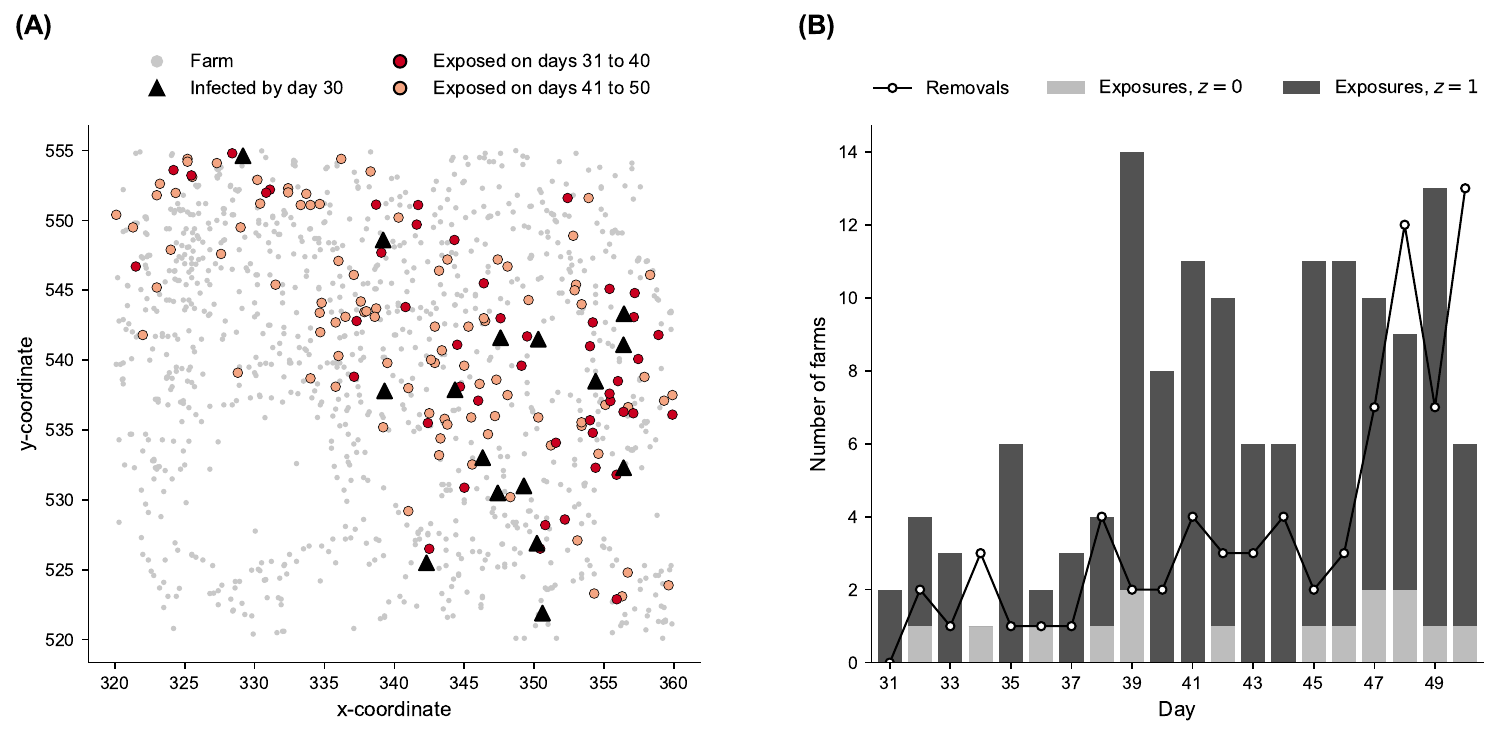}
    \caption{
    Observed FMD outbreak. (A) Farm locations, with farms infected by
    day 30 (triangles) and farms exposed on days 31 to 40 (dark circles)
    and 41 to 50 (light circles). (B) Daily new exposures by $z_i$ and
    daily removals.
    }
    \label{fig:fmd_data}
\end{figure}

The FMD data, compiled by \citet{deeth2013latent}, cover 1,177 cattle and
sheep farms in a subregion of Cumbria in north-west England. For each
farm, the records give its Cartesian coordinates, its infection and
culling dates, and its numbers of cattle and sheep. Following
\citet{ward2025framework} and \citet{mao2026memory}, we restricted the analysis
to days 30 to 50 of the epidemic (8 to 28 March 2001). The state of the
outbreak on day 30 served as the initial condition, with 6 infectious,
9 exposed, 1 removed, and 1,161 susceptible farms, and the model was
fitted to the 140 new exposures that occurred on days 31 to 50
(Figure~\ref{fig:fmd_data}).

Farm size entered the model through a binary covariate $z_i$. Using the
criteria of \citet{deeth2013latent}, a farm with more than 50 cattle or
more than 200 sheep was assigned $z_i=1$ (735 farms), and each of the
remaining 442 farms was assigned $z_i=0$.

Transmission was described by an SEIR individual-level model, which
adds an exposed (E) state to the SIR model of
Section~\ref{sec:model}. A newly infected farm first enters the exposed
state, in which it cannot yet transmit infection.

As in \citet{ward2025framework}, the exposed period was fixed at 5 days.
A susceptible farm $i$ becomes exposed at
time $t+1$ with probability
\begin{equation}
  P(i,t)
  = 1 - \exp\!\left[
  -(\alpha_0 + \alpha_1 z_i)
  \sum_{j \in I(t)} (d_{ij} + 1)^{-\beta}
  \right],
  \label{eq:fmd_exposure}
\end{equation}
where $I(t)$ is the set of infectious farms at time $t$, and
$\vartheta=(\alpha_0,\alpha_1,\beta)$. 
To simulate epidemics, the infectious period of each farm infected
during a simulation was drawn from the empirical distribution of
infectious periods among farms whose infectious period began on days 31
to 50. This distribution places probability 0.16 on 3 days and 0.84 on
4 days, with mean $\mathbb{E}[\mathrm{IP}]=3.84$ days. 

The prior followed the construction of
Section~\ref{sec:simulation_setup}, with
$\beta \sim \mathrm{Gamma}(6,1/4)$ and
$R_0 \sim \mathrm{Gamma}(10,1/4)$. To accommodate the covariate, we
further assigned $r=\alpha_0/(\alpha_0+\alpha_1) \sim \mathrm{U}(0,1)$,
the susceptibility of a farm with $z_i=0$ relative to a farm with
$z_i=1$. The reproduction number was defined as
\begin{equation}
  R_0
  = \mathbb{E}[\mathrm{IP}]\,
  \frac{1}{|\mathcal{I}_0|}
  \sum_{j \in \mathcal{I}_0}
  \sum_{i \in \mathcal{S}_0}
  (\alpha_0 + \alpha_1 z_i)(d_{ij} + 1)^{-\beta},
  \label{eq:fmd_r0}
\end{equation}
where $\mathcal{I}_0$ and $\mathcal{S}_0$ are the sets of infectious and
susceptible farms on day 30.  
Figure~S7 in the
Supplementary Information shows the prior and prior
predictive distributions.

\subsection{Inference and Diagnostics}
\label{sec:fmd_inference}

Within the analysis window, exposure dates follow from the dates of
infectiousness and the fixed exposed period, and removal dates are
recorded. The set $I(t)$ is therefore known on every day, and the
likelihood
\begin{equation}
  f(D \mid \vartheta)
  = \prod_{t}
  \Bigg[ \prod_{i \in E(t+1) \setminus E(t)} P(i,t) \Bigg]
  \Bigg[ \prod_{i \in S(t+1)} \big\{1 - P(i,t)\big\} \Bigg]
  \label{eq:fmd_lik}
\end{equation}
can be evaluated exactly, where $E(t)$ and $S(t)$ denote the sets of
exposed and susceptible farms at time $t$. MCMC therefore requires no data augmentation, targets the posterior under the model directly, and provides a high-accuracy reference. We used an adaptive random-walk Metropolis--Hastings sampler
on $(\log\alpha_0, \log\alpha_1, \log\beta)$ with 4 chains of 100,000
iterations, discarding the first 20,000 as burn-in. 

The CNN and GNN embeddings followed
Section~\ref{sec:npe_embeddings}. The CNN received the daily numbers of
new exposures among farms with $z_i=0$ and with $z_i=1$ and the daily
number of removals. The GNN received, for each farm, its exposure and
removal times within the window, its covariate $z_i$, and its coordinates. Each estimator was trained on 20,000 epidemics simulated
from the prior. To assess training variability, each architecture was
trained five times with different random seeds, and we pooled 3,000
posterior draws from each trained network. 

Without true parameter values, we used checks that condition on the
observed data $\mathcal{D}^{\mathrm{obs}}$. L-C2ST \citep{linhart2023lc2st}
tests whether the NPE posterior at $\mathcal{D}^{\mathrm{obs}}$ equals the
true posterior. The flow in Eq.~\eqref{eq:flow_density} maps each
parameter value to a base-space value
$z=T_\phi^{-1}\{\vartheta \mid h_\psi(\mathcal{D}^{\mathrm{obs}})\}$. If
the NPE posterior is exact, $z$ follows the standard normal base
distribution $\pi_0$ independently of $\mathcal{D}^{\mathrm{obs}}$ when
$(\vartheta,\mathcal{D}^{\mathrm{obs}})$ is drawn from the joint
distribution of parameters and data. Using a reference set of 10,000
simulations $(\vartheta^{(n)},\mathcal{D}^{\mathrm{obs},(n)})$ from the
prior, separate from the training set, we trained a classifier $c$ to
separate $\{T_\phi^{-1}(\vartheta^{(n)} \mid h^{(n)}),h^{(n)}\}$ from
$\{u^{(n)},h^{(n)}\}$, where $h^{(n)}=h_\psi(\mathcal{D}^{\mathrm{obs},(n)})$
and $u^{(n)}\sim\mathcal{N}(0,I_3)$. Under an exact posterior, the two
class distributions coincide, and the Bayes-optimal classifier assigns
probability $c=0.5$. The statistic
\begin{equation}
  \Lambda(\mathcal{D}^{\mathrm{obs}})
  = \frac{1}{N_u}\sum_{m=1}^{N_u}
    \big[c\{u_m,h_\psi(\mathcal{D}^{\mathrm{obs}})\}-0.5\big]^2,
  \qquad u_m\sim\mathcal{N}(0,I_3),
  \label{eq:fmd_lc2st}
\end{equation}
with $N_u=10{,}000$, measures the departure from the true posterior at
$\mathcal{D}^{\mathrm{obs}}$. Null statistics
$\Lambda_1,\ldots,\Lambda_{50}$ were obtained from 50 classifiers trained
with the base-space values replaced by standard normal draws, and
\begin{equation}
  p = \frac{1+\sum_{b=1}^{50}\mathbf{1}\{\Lambda_b \ge \Lambda(\mathcal{D}^{\mathrm{obs}})\}}{51},
  \label{eq:fmd_lc2st_p}
\end{equation}
so the smallest attainable $p$-value is $1/51$.

An L-C2ST rejection, or a difference between NPE and MCMC at $\mathcal{D}^{\mathrm{obs}}$, can
arise from the approximation error of NPE, which would also appear on
data simulated from the model, or from features of the observed outbreak
that the simulator does not reproduce. To help distinguish the two, we
generated 50 posterior predictive datasets, each simulated from a
parameter draw from the MCMC posterior. Each of these local datasets was analyzed with MCMC and with
all ten trained networks. On each local dataset we computed $\Lambda$ and the
shift of each NPE posterior median from the MCMC posterior median on
the log scale, and compared them with their values at $\mathcal{D}^{\mathrm{obs}}$.

Posterior
predictive checks compared the observed cumulative exposures and the
mean distance from each newly exposed farm to the nearest farm
infectious on the preceding day, which reflects the spatial range of
transmission. In Section~S7 of the
Supplementary Information, we also used the maximum mean discrepancy (MMD) and the Mahalanobis distance between the embedding of the observed data and
those of held-out simulations to test whether the observed data lie
within the range of the training data
\citep{schmitt2023misspecification}.

\subsection{Results}
\label{sec:fmd_results}

\begin{figure}[h!]
    \centering
    \includegraphics[width=\linewidth]{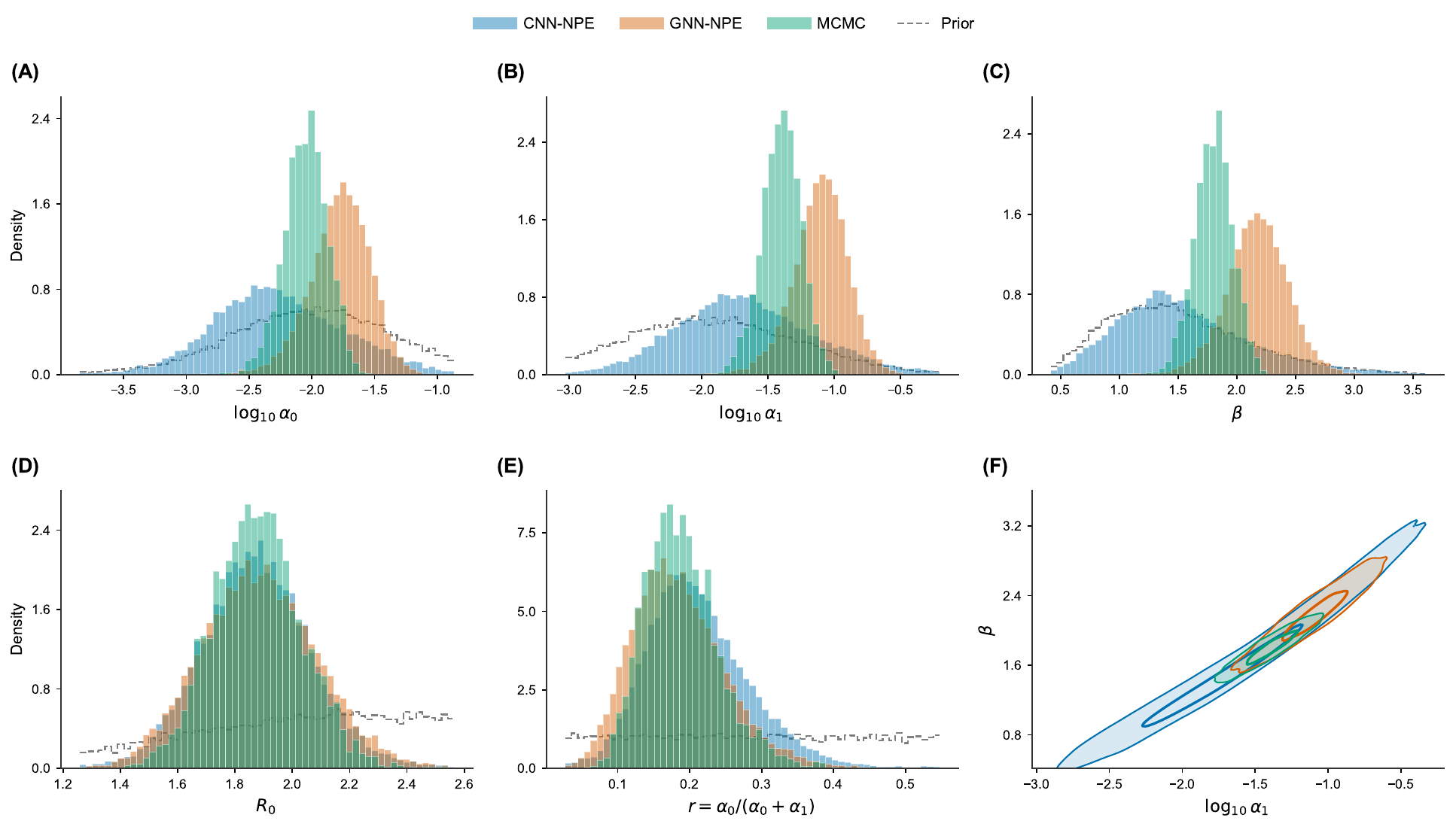}
    \caption{
    Posteriors for the observed outbreak. (A to E) Marginal posteriors,
    with the prior as a dashed line. (F) Joint posterior of
    $\log_{10}\alpha_1$ and $\beta$, with contours enclosing 50\% and
    95\% of the posterior mass.
    }
    \label{fig:fmd_posterior}
\end{figure}

The MCMC chains mixed well, with rank-normalized $\hat{R}$ at most
1.0003 and bulk effective sample sizes above 24,000 for all parameters
(Table~S9 in the Supplementary Information).
Table~\ref{tab:fmd_posterior} and Figure~\ref{fig:fmd_posterior}A to E
summarize the posteriors.
Under MCMC, $R_0$ was 1.87 (95\% credible interval 1.59 to 2.18) and $r$
was 0.18 (0.10 to 0.31), so farms with $z_i=0$ were about one fifth as
susceptible as farms with $z_i=1$, in line with the observed attack
rates of 3.4\% and 17.4\% (Figure~\ref{fig:fmd_data}). CNN-NPE and
GNN-NPE gave posteriors for $R_0$ and $r$ close to the MCMC posterior.
Their posteriors for $\alpha_0$, $\alpha_1$, and $\beta$ differed from the MCMC posterior. The GNN-NPE medians lay above and the CNN-NPE medians below the MCMC medians, and the CNN-NPE posteriors were much wider.

\begin{table}[h!]
\centering
\caption{Posterior medians and 95\% credible intervals for the observed
outbreak. CNN-NPE and GNN-NPE summaries pool 3,000 posterior draws from
each of five trained networks. $R_0$ and $r=\alpha_0/(\alpha_0+\alpha_1)$
are computed from each posterior draw using Eq.~\eqref{eq:fmd_r0}. }
\label{tab:fmd_posterior}
\small
\setlength{\tabcolsep}{4pt}
\begin{tabular}{lccccc}
\toprule
Method & $\alpha_0$ & $\alpha_1$ & $\beta$ & $R_0$ & $r$ \\
\midrule
MCMC & 0.0091 [0.0040, 0.018] & 0.040 [0.019, 0.076] & 1.81 [1.48, 2.11] & 1.87 [1.59, 2.18] & 0.18 [0.10, 0.31] \\
CNN-NPE & 0.0051 [0.00052, 0.065] & 0.020 [0.0022, 0.28] & 1.48 [0.63, 2.99] & 1.87 [1.49, 2.28] & 0.20 [0.09, 0.38] \\
GNN-NPE & 0.017 [0.0051, 0.045] & 0.080 [0.027, 0.19] & 2.18 [1.63, 2.68] & 1.88 [1.50, 2.30] & 0.18 [0.08, 0.33] \\
\bottomrule
\end{tabular}
\end{table}

Figure~\ref{fig:fmd_posterior}F explains this pattern. The joint
posterior of $\alpha_1$ and $\beta$ lies along a narrow increasing ridge,
and the MCMC posterior correlation between $\log\alpha_1$ and $\beta$
is 0.93. A larger $\beta$ makes infectious pressure decay faster with
distance, so a larger transmission rate is needed to produce the same
number of exposures. Points along the ridge give similar values of
$R_0$, which the numbers of exposures determine well. The position on
the ridge depends on how far transmission reaches, and only the spatial
pattern of exposures carries this information. The three posteriors lie
on the same ridge. The MCMC posterior is concentrated near its middle,
the GNN-NPE posterior lies further up the ridge, and the CNN-NPE
posterior extends along most of it.

\begin{figure}[h!]
    \centering
    \includegraphics[width=\linewidth]{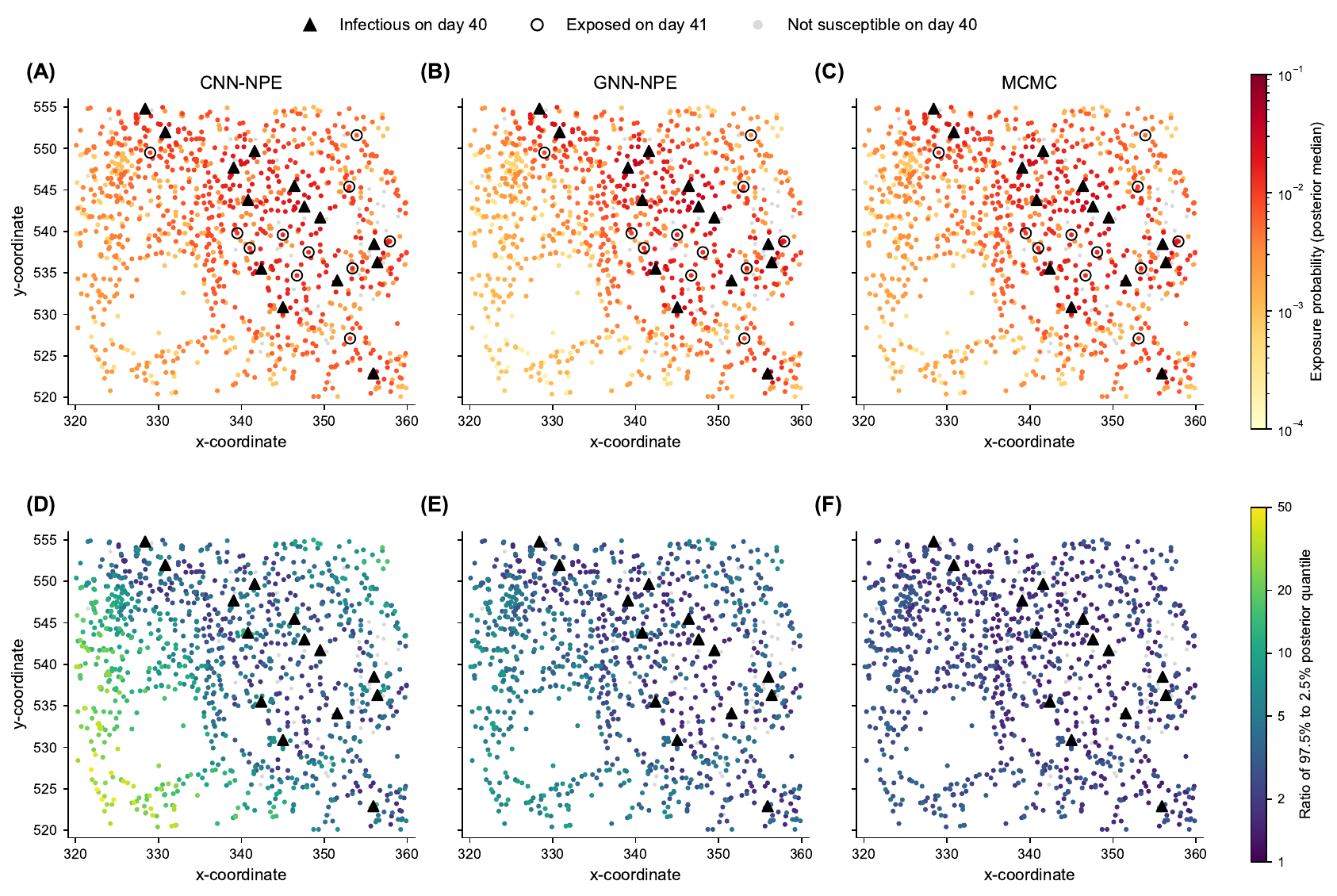}
    \caption{
    Posterior risk of exposure on day 41 given the farms infectious on
    day 40 (triangles). (A to C) Posterior median of the exposure
    probability of each susceptible farm, with the 11 farms exposed on
    day 41 circled. (D to F) Ratio of the 97.5\% to the 2.5\% posterior
    quantile of this probability. Results for days 35 and 45 were
    similar.
    }
    \label{fig:fmd_risk}
\end{figure}

Figure~\ref{fig:fmd_risk} shows how these differences appear in space.
Using the 14 farms infectious on day 40, we computed the probability
that each susceptible farm was exposed on day 41 under each posterior
draw. The expected number of exposures was almost the same for the
three methods, but the uncertainty in farm-level risk differed. The
median ratio of the 97.5\% to the 2.5\% posterior quantile was 2.0
under MCMC, 3.6 under GNN-NPE and 5.1 under CNN-NPE. Incidence counts
carry little information on where infections occur, so CNN-NPE cannot
determine how far transmission reaches, and GNN-NPE, which uses the
locations of exposures, is closer to the MCMC posterior.

\begin{figure}[h!]
    \centering
    \includegraphics[width=\linewidth]{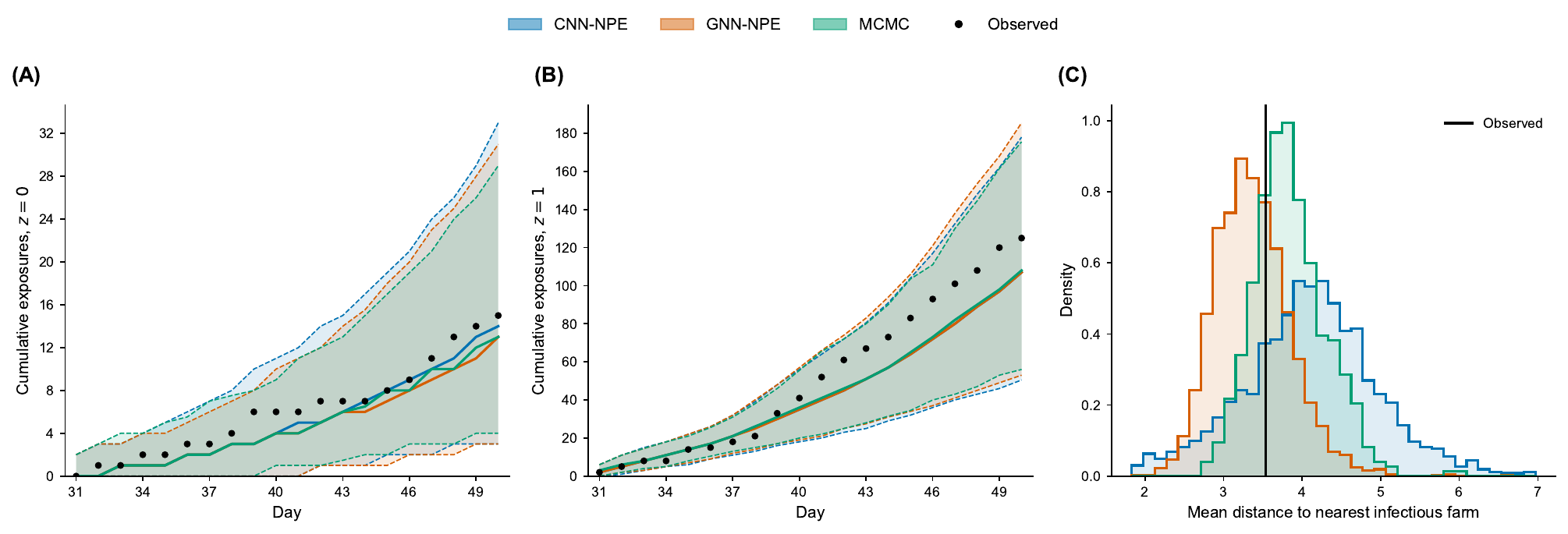}
    \caption{
    Posterior predictive checks. (A, B) Cumulative new exposures for
    $z_i=0$ and $z_i=1$, with predictive medians (solid), 95\%
    predictive intervals (dashed and shaded) and observed values
    (points). (C) Predictive distributions of the mean distance from each
    newly exposed farm to the nearest farm infectious on the preceding
    day, with the observed value as a vertical line.
    }
    \label{fig:fmd_ppc}
\end{figure}

Posterior predictive checks showed no clear lack of fit
(Figure~\ref{fig:fmd_ppc}). The observed cumulative exposures stayed
within the 95\% predictive intervals of all three methods. For the mean
transmission distance, the observed value lay within each predictive
interval. GNN-NPE predicted shorter distances (median 3.34, against 3.81
under MCMC, 4.11 under CNN-NPE and 3.54 observed), consistent with its
larger $\beta$.

\begin{figure}[h!]
    \centering
    \includegraphics[width=\linewidth]{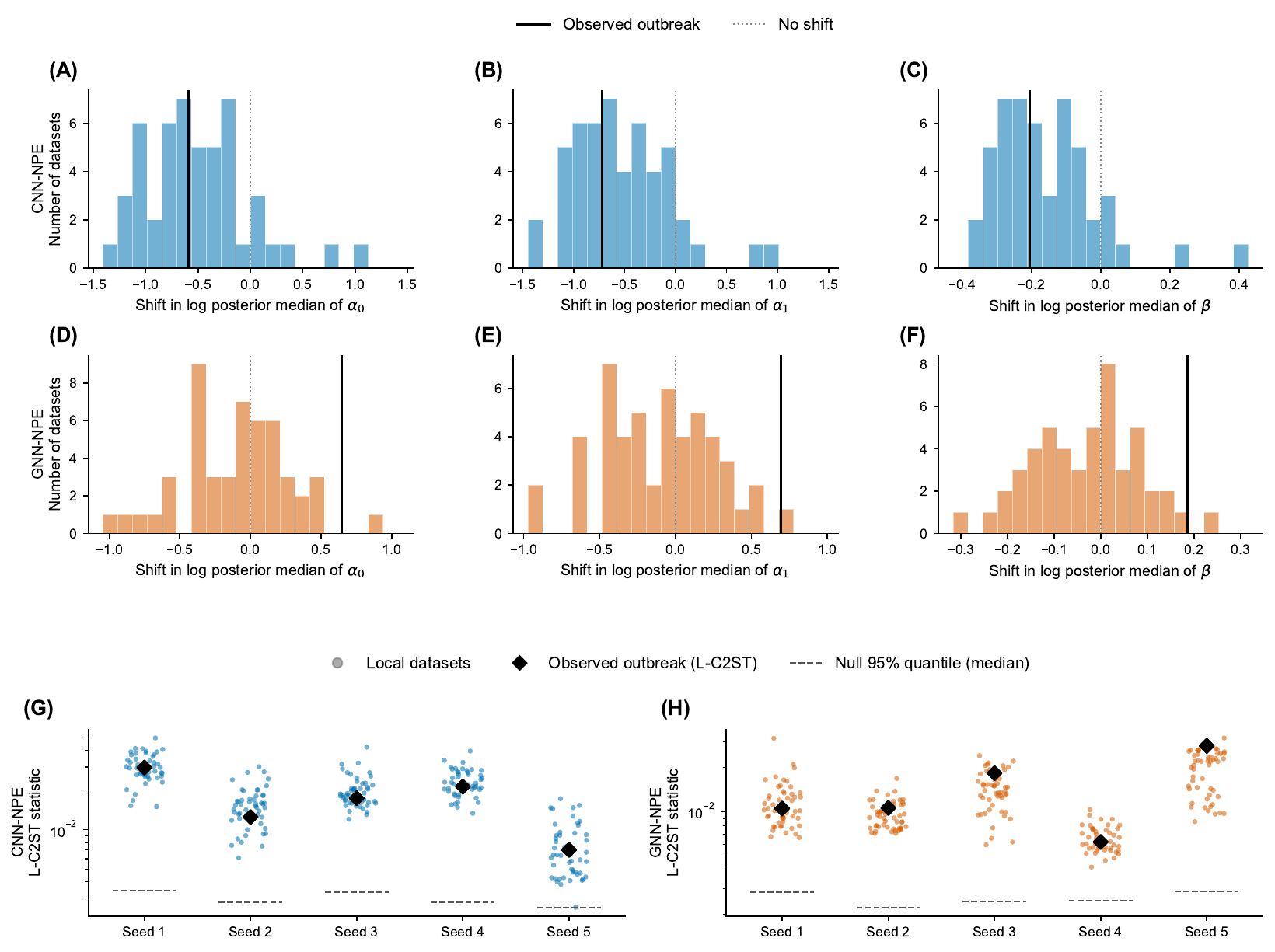}
    \caption{
    Local check around the observed data. (A to C) CNN-NPE and (D to F)
    GNN-NPE shift of the log posterior median from the log MCMC
    posterior median on 50 posterior predictive datasets simulated from
    MCMC posterior draws,
    averaged over the five trained networks, with the shift at the
    observed outbreak as a solid line. (G) CNN-NPE and (H) GNN-NPE
    L-C2ST statistic of each trained network on the local datasets
    (points) and at the observed outbreak (diamonds), with the median
    over local datasets of the 95\% quantile of the null distribution
    (dashed).
    }
    \label{fig:fmd_local}
\end{figure}

L-C2ST rejected the hypothesis of an exact posterior at the observed data for all ten trained networks
($p\le 0.039$). Rejection was also widespread on the local datasets
(94\% to 100\% of datasets for each network), and the statistic at the
observed data lay between the 28th and 98th percentiles of the local
statistics (Figure~\ref{fig:fmd_local}G and H). This indicates that the
rejection at the observed data arises from approximation error of NPE
that is common in the posterior predictive neighborhood of the observed
data. The local datasets also show where the differences between NPE and MCMC
come from (Table~\ref{tab:fmd_local} and
Figure~\ref{fig:fmd_local}A to F). On these datasets, the CNN-NPE
posterior medians of $\alpha_0$ and $\alpha_1$ lay about 0.5 below the
MCMC medians on the log scale, and the CNN-NPE 95\% credible intervals were about three times as wide as those of MCMC. The CNN-NPE shifts at the
observed data were close to the center of their local distributions, so
the CNN-NPE posterior at the observed data is typical of this
estimator. GNN-NPE showed no systematic shift on the local datasets,
with a standard deviation of about 0.4 for $\alpha_0$ and $\alpha_1$.
Its shifts at the observed data (0.65 and 0.69 for $\alpha_0$ and
$\alpha_1$) were reached or exceeded on only 1 of the 50 local datasets. Because $\alpha_0$, $\alpha_1$, and $\beta$ move together along the ridge, these
shifts form a single departure along it, which is atypical of the
approximation error of GNN-NPE under the model. This suggests that the
observed outbreak contains features that the simulator does not
reproduce, to which GNN-NPE responds differently from the likelihood.

\begin{table}[h!]
\centering
\caption{Local check on 50 posterior predictive datasets simulated from MCMC posterior draws. All quantities are on the log scale. Error is the posterior median minus the true value, given as mean (SD) over datasets and trained networks. Width is the mean width of the 95\% credible interval. Shift is the NPE posterior median minus the MCMC posterior median on the same dataset, given as mean (SD). Observed shift is the same quantity at the observed outbreak, averaged over the five trained networks. The last column counts local datasets whose network-averaged shift is at least the observed shift.}
\label{tab:fmd_local}
\small
\setlength{\tabcolsep}{4pt}
\begin{tabular}{llccccc}
\toprule
Method & Parameter & Error & Width & Shift & Observed shift & Datasets $\geq$ observed \\
\midrule
MCMC & $\alpha_0$ & $-0.07$ (0.39) & 1.69 &  &  &  \\
 & $\alpha_1$ & $-0.16$ (0.44) & 1.42 &  &  &  \\
 & $\beta$ & $-0.04$ (0.11) & 0.40 &  &  &  \\
CNN-NPE & $\alpha_0$ & $-0.58$ (0.41) & 4.86 & $-0.51$ (0.51) & $-0.59$ & 25/50 \\
 & $\alpha_1$ & $-0.70$ (0.44) & 4.82 & $-0.53$ (0.50) & $-0.72$ & 31/50 \\
 & $\beta$ & $-0.19$ (0.11) & 1.57 & $-0.16$ (0.15) & $-0.21$ & 29/50 \\
GNN-NPE & $\alpha_0$ & $-0.17$ (0.53) & 2.54 & $-0.10$ (0.40) & 0.65 & 1/50 \\
 & $\alpha_1$ & $-0.27$ (0.46) & 2.26 & $-0.11$ (0.38) & 0.69 & 1/50 \\
 & $\beta$ & $-0.07$ (0.13) & 0.70 & $-0.03$ (0.12) & 0.19 & 1/50 \\
\bottomrule
\end{tabular}
\end{table}

\section{Discussion}
\label{sec:discussion}

This paper developed and evaluated neural posterior estimation for
Bayesian inference in spatial individual-level epidemic models, using a
simulation study under three observation scenarios and an analysis of
the observed 2001 UK FMD outbreak. Posterior accuracy for the spatial
transmission parameters depended on the embedding architecture, and the
graph embedding improved inference for the transmission intensity $\alpha$ and especially for the
distance-decay parameter $\beta$. When its chains converged,
likelihood-based MCMC gave the most accurate posteriors, and GNN-NPE was
the closer of the two NPE variants to MCMC at a small fraction of the per-epidemic cost of MCMC.

The comparison between CNN-NPE and GNN-NPE shows the role of spatial
information in posterior estimation. The CNN embedding summarizes each
epidemic through the population-level incidence curve, capturing
temporal growth, peak timing, and epidemic size, and discards where
infections occur. The GNN embedding retains individual locations and
local neighborhood structure, so the posterior estimator can learn from
spatial clustering and the range of transmission. This explains why the
largest gains appeared for $\alpha$ and $\beta$, which control the
intensity and spatial decay of transmission, and why the CNN embedding
performed as well as or better than the GNN embedding for $\gamma$ and
$\rho$, whose information lies mainly in the timing and number of
observed infections. 
Together, these results suggest that the effectiveness of an embedding depends on how well the information it preserves aligns with the inferential target.

MCMC uses the likelihood of the complete or augmented data and, when the
sampler mixes well, targets the posterior defined by the model and
prior. Among the epidemics with converged chains, MCMC gave the most
accurate posteriors in all three scenarios, with 95\% credible intervals
for $\alpha$ and $\beta$ narrower than those of GNN-NPE. With data
augmentation, the sampler also returns posterior draws of the unobserved
removal or infection times of each individual, so the inferred epidemic
history can be interpreted at the individual level, for example through
the posterior probability that a given individual was infected but not
reported.

The advantage of NPE lies in its computational cost. Once trained, a
network returns a posterior for a new epidemic at a small fraction of
the cost of an MCMC run, so the one-off cost of simulation and training
is shared across all datasets analyzed with the same estimator. The cost
of NPE also does not depend on the mixing of latent variables.
Data-augmented MCMC mixed slowly in Partial at low observation
probabilities, where the number and timing of hidden infections were
weakly constrained by the data, and a substantial fraction of runs did
not meet the convergence criterion despite long chains. The price of
this speed is that NPE returns a learned approximation, whose quality
depends on the training data, the capacity of the neural density
estimator, and how much of the information in the observation the
embedding retains. NPE posteriors therefore require validation through
simulation-based calibration \citep{talts2018validating}, posterior
predictive checks, or comparison with MCMC where the likelihood is
tractable.

NPE learns the posterior under the simulator and prior used to generate
the training data. When the observed data differ from what the simulator
produces, the estimator extrapolates beyond its training regime, and its
posterior can differ from the likelihood-based posterior in ways that
calibration checks on simulated data do not reveal. Methods have been
proposed to detect such misspecification in amortized inference and to
make NPE robust to it \citep{schmitt2023misspecification,ward2022robust}.
In the FMD analysis, the embedding checks did not flag the observed
outbreak, yet the GNN-NPE posterior at the observed data departed from
the likelihood-based posterior more than on data simulated from the
model. One possible source of this difference is a change in the control
response during the outbreak. Within the analysis window, the proportion of farms culled within 3 days
of the onset of infectiousness rose from 8\% for farms that became
infectious on days 31 to 40 to 19\% for those that became infectious on
days 41 to 50, whereas the simulator drew infectious periods from a
single distribution over the analysis window.

A natural extension is to apply the same framework to broader classes of
ILMs. The models considered here used relatively simple spatial SIR and
SEIR structures, but ILMs can represent much richer transmission
mechanisms. NPE could be extended to behavior-change ILMs, where
individual susceptibility is modulated by a population-level alarm
function that captures perceived epidemic risk over time
\citep{ward2025framework,mao2026memory}; geographically dependent ILMs, where transmission risk
depends on spatially varying risk factors and unobserved geographical
structure \citep{Mahsin2022Geographically}; multistrain ILMs, where
multiple pathogen strains circulate and interact within the same
population \citep{romanescu2016modeling}; network-based ILMs, where
transmission is mediated by observed or latent contact networks
\citep{almutiry2021contact}; continuous-time ILMs, where infection and
removal events occur in continuous time \citep{almutiry2021epiilmct};
and directional spatial ILMs, where transmission may depend on movement,
wind, commuting, or other directional processes
\citep{peitsch2025directional}. These extensions are especially relevant
because richer ILMs typically make likelihood evaluation and
data-augmented MCMC more difficult, while forward simulation often
remains straightforward.

Further work is also needed on model comparison. This paper focused on
parameter inference under a specified model. In practice, analysts often
need to compare alternative transmission kernels, observation models, or
intervention mechanisms. Standard Bayesian model comparison relies on
marginal likelihoods, which are generally unavailable when the
likelihood is intractable. Amortized neural methods can approximate
posterior model probabilities directly from simulations
\citep{radev2021amortized,chatha2026stochastic}, and simulation-based model criticism can
assess whether a fitted model reproduces the observed data
\citep{ward2022robust,schmitt2023misspecification}. Another direction is
to use NPE as a fast screening tool for candidate models, followed by
likelihood-based analysis of the most plausible specifications.

In conclusion, NPE provides a fast approximation to Bayesian inference
in spatial ILMs, which is useful when inference must be repeated across
many datasets or when data-augmented MCMC mixes slowly. Spatially
structured embeddings improve inference for spatial transmission
parameters and bring NPE closer to the likelihood-based posterior.
Combined with calibration checks on simulated data and local diagnostics
at the observed data, NPE can extend the practical scope of Bayesian
inference for individual-level epidemic models.

\section*{Acknowledgments}
This research was supported by the Natural Sciences and Engineering
Research Council of Canada (NSERC) Discovery Grant program
(RGPIN/03292-2022) awarded to R. D..
Y.M. was supported by a Postdoctoral Fellowship from the Pacific Institute for the Mathematical Sciences (PIMS).
The funders had no role in study design, data collection and analysis, decision to publish, or preparation of the manuscript.
\section*{Data and Code Availability}
The data and code used in this study is available at
\url{https://github.com/YichengMao98/NPE-ILM}.

\section*{Author Contribution}

Yicheng Mao: Conceptualization, Formal analysis, Investigation, Methodology, Software, Validation, Visualization, Writing – original draft, Writing – review \& editing\\
\noindent Rob Deardon: Conceptualization, Methodology, Supervision, Writing – review \& editing

\bibliography{ref}

\end{document}